\documentclass[12pt]{article}
\usepackage{epsfig,cite}
\usepackage {amsmath}
\usepackage{amssymb}
\usepackage{ulem}
\include{epsf}
\newcount\timecount
\newcount\hours \newcount\minutes  \newcount\temp \newcount\pmhours
\hours = \time \divide\hours by 60 \temp = \hours \multiply\temp by
60 \minutes = \time \advance\minutes by -\temp
\def\hour{\the\hours}
\def\minute{\ifnum\minutes<10 0\the\minutes
            \else\the\minutes\fi}
\def\clock{
\ifnum\hours=0 12:\minute\ AM \else\ifnum\hours<12 \hour:\minute\ AM
      \else\ifnum\hours=12 12:\minute\ PM
            \else\ifnum\hours>12
                 \pmhours=\hours
                 \advance\pmhours by -12
                 \the\pmhours:\minute\ PM
                 \fi
            \fi
      \fi
\fi }

\def\monthname{\relax\ifcase\month 0/\or January\or February\or
   March\or April\or May\or June\or July\or August\or September\or
   October\or November\or December\else\number\month/\fi}

\def\bold#1{\setbox0=\hbox{$#1$}%
     \kern-.025em\copy0\kern-\wd0
     \kern.05em\copy0\kern-\wd0
     \kern-.025em\raise.0433em\box0 }

\def\beq{\begin{equation}}
\def\eeq{\end{equation}}

\def\ga{\mathrel{\raise.3ex\hbox{$>$\kern-.75em\lower1ex\hbox{$\sim$}}}}
\def\la{\mathrel{\raise.3ex\hbox{$<$\kern-.75em\lower1ex\hbox{$\sim$}}}}
\def\gev{{\rm \, Ge\kern-0.125em V}}
\def\tev{{\rm \, Te\kern-0.125em V}}
\def\gyr{{\rm \, G\kern-0.125em yr}}

\def\gappeq{\mathrel{\rlap {\raise.5ex\hbox{$>$}}
{\lower.5ex\hbox{$\sim$}}}}
\def\lappeq{\mathrel{\rlap{\raise.5ex\hbox{$<$}}
{\lower.5ex\hbox{$\sim$}}}}
\def\Toprel#1\over#2{\mathrel{\mathop{#2}\limits^{#1}}}

\def\m12{m_{1\!/2}}
\def\bea{\begin{eqnarray}}
\def\eea{\end{eqnarray}}

\begin{document}
\begin{titlepage}
\pagestyle{empty} \baselineskip=21pt
%\rightline{\tt hep-ph/yymmnnn}
\rightline{CERN-PH-TH/2008-088} \vskip 0.2in
\begin{center}
{\large{\bf Probes of Lorentz Violation in Neutrino Propagation}}
\end{center}
\begin{center}
\vskip 0.2in {\bf John~Ellis}$^1$, {\bf Nicholas~Harries}$^{1,2}$,
{\bf Anselmo~Meregaglia}$^{3}$, {\bf Andr\' e~Rubbia}$^{4}$ and {\bf
Alexander~S.~Sakharov}$^{1,4}$ \vskip 0.1in

{\it
$^1${TH Division, PH Department, CERN, CH-1211 Geneva 23, Switzerland}\\
$^2${Theoretical Physics, University of Oxford, 1Keble Road, Oxford, UK}\\
$^3$ IPHC, Universit, Louis Pasteur, CNRS/IN2P3, Strasbourg, France\\
$^4$ Swiss Institute of Technology ETH-Z\"urich, CH-8093 Z\"urich,
Switzerland\\
}

\vskip 0.2in {\bf Abstract}
\end{center}
\baselineskip=18pt \noindent
%%%%%%%%%%%%%%%%%%%%%%%%%%%%%%%%%%%%%%%%%%%%%%%%%%%%%%%%%%%%%%%%%%%%%

It has been suggested that the interactions of energetic particles
with the foamy structure of space-time thought to be generated by
quantum-gravitational (QG) effects might violate Lorentz invariance,
so that they do not propagate at a universal speed of light.  We
consider the limits that may be set on a linear or quadratic
violation of Lorentz invariance in the propagation of energetic
neutrinos, $v/c = [1 \pm (E/M_{\nu QG1})]$ or $[1 \pm (E/M_{\nu
QG2})^2]$, using data from supernova explosions and the OPERA
long-baseline neutrino experiment. Using the SN1987a neutrino data
from the Kamioka II, IMB and Baksan experiments, we set the limits
$M_{\nu QG1} > 2.7 (2.5) \times 10^{10}$~GeV for subluminal
(superluminal) propagation, respectively, and $M_{\nu QG2} > 4.6
(4.1) \times 10^{4}$~GeV at the 95~\% confidence level. A future
galactic supernova at a distance of 10~kpc would have sensitivity to
$M_{\nu QG1} > 2 (4) \times 10^{11}$~GeV for subluminal
(superluminal) propagation, respectively, and $M_{\nu QG2} > 2 (4)
\times 10^{5}$~GeV. With the current CNGS extraction spill length of
10.5~$\mu$s and with standard clock synchronization techniques,  the
sensitivity of the OPERA experiment would reach $M_{\nu QG1} \sim
7\times 10^{5}$~GeV ($M_{\nu QG2} \sim 8 \times 10^{3}$~GeV) after 5
years of nominal running. If the time structure of the SPS RF
bunches within the extracted CNGS spills could be exploited, these
figures would be significantly improved to $M_{\nu QG1}\sim 5 \times
10^{7}$~GeV ($M_{\nu QG2} \sim 4 \times 10^{4}$~GeV).
These results can be improved further if similar time resolution can
be achieved with neutrino events occurring in the rock upstream of
the OPERA detector: we find potential sensitivities to $M_{\nu
QG1}\sim 4 \times 10^{8}$~GeV  and $M_{\nu QG2} \sim 7 \times
10^{5}$~GeV.

\vspace*{0.2cm}

%%%%%%%%%%%%%%%%%%%%%%%%%%%%%%%%%%%%%%%%%%%%%%%%%%%%%%%%%%%%%%%%%%%%%%
\vfill \leftline{CERN-PH-TH/2008-088} \leftline{April 2008}
\end{titlepage}
%\baselineskip=18pt
%%%%%%%%%%%%%%%%%%%%%%%%%%%%%%%%%%%%%%%%%%%%%%%%%%%%%%%%%%%%%%%%%%%%%%

\section{Introduction}

Neutrinos from astrophysical sources  and long-baseline experiments
are powerful probes of potential new physics. They have already been
used to discover and measure the novel phenomena of neutrino
oscillations, thereby establishing that neutrino have masses
\cite{Strumia:2006db, Totsuka:1991dm}. Long-baseline neutrino
experiments have also been used to set limits on quantum decoherence
effects that might be induced by foamy fluctuations in the
space-time background in some models of quantum gravity
(QG)~\cite{Barenboim:2006xt, CNGS-T2K, Morgan:2004vv,
Hooper:2004xr}. It has also been suggested that the space-time foam
due to QG fluctuations might cause energetic particles to propagate
at speeds different from the velocity of light, which would be
approached only by low-energy massless particles~\cite{foam,
gambini}. Any deviation from the velocity of light at high energies
might be either linear or quadratic, $\delta v/c = (E/M_{QG1})$ or
$(E/M_{QG2})^2$, and might be either subluminal or superluminal.
Such effects are, in principle, easily distinguishable from the
effects of neutrino masses, since they depend differently on the
energy $E$.

There have been many tests of such Lorentz-violating effects on
photon propagation from distant astrophysical objects such as
gamma-ray bursters~\cite{amellis}, pulsars~\cite{pulsar} and active
galactic nuclei~\cite{Albert:2007qk}. These tests have looked for
delays in the arrival times of energetic photons relative to
low-energy photons, and their sensitivities improve with the
distance of the source, the energies of the photons, the accuracy
with which the arrival times of photons can be measured, and the
fineness of the time structure of emissions at the astrophysical
source. The sensitivities of these tests has reached $M_{\gamma QG1}
\sim 2 \times 10^{17}$~GeV and $M_{\gamma QG2} \sim 4 \times
10^{10}$~GeV for linear and quadratic violations of Lorentz
invariance, respectively.

At least one QG model of space-time foam~\cite{equiv,refract_last}
suggests that Lorentz violation should be present only for particles
without conserved internal quantum numbers, such as photons, and
should be absent for particles with electric charges, such as
electrons. Indeed, astrophysical data have been used to set very
stringent limits on any Lorentz violation in electron propagation.
However, these arguments do not apply to neutrinos, since they are
known to oscillate, implying that lepton flavour quantum numbers are
not conserved. Moreover, neutrinos are often thought to be Majorana
particles, implying that the overall lepton number is also not
conserved, in which case QG effects might also be present in
neutrino propagation~\cite{LeptonNumber}. It therefore becomes
interesting to study experimentally the possibility of Lorentz
violation in neutrino propagation.

Experimental probes of Lorentz violation in neutrino propagation are
hindered by the relative paucity of neutrino data from distant
astrophysical sources, and require the observation of narrow time
structures in neutrino emissions. However, there has been one
pioneering experimental study of possible Lorentz violation using
the long-baseline MINOS experiment exposed to the NuMI neutrino beam
from Fermilab, which found a range of neutrino velocities $-2.4
\times 10^{-5} < (v - c)/c < 12.6 \times 10^{-5}$ allowed at the
99\% C.L.~\cite{Rebel:2008th}. Assuming an average neutrino energy
of 3~GeV, and allowing for either linear or quadratic Lorentz
violation: $v/c = [1 \pm (E/M_{\nu QG1})]$ or $[1 \pm (E/M_{\nu
QG2})^2]$, the MINOS result~\cite{Rebel:2008th} corresponds in the
case of linear Lorentz violation to $M_{\nu QG1}
> 1 (4) \times 10^5$~GeV in the case of subluminal (superluminal)
propagation, and in the case of quadratic Lorentz violation to
$M_{\nu QG2} > 600 (250)$~GeV.

In this paper we first establish limits on Lorentz violation using
neutrino data from supernova 1987a, using data from the Kamioka II
(KII) \cite{k2sn1987a}, Irvine-Michigan-Brookhaven (IMB)
\cite{imbsn1987a} and Baksan detectors \cite{baksan1987a}. We find
$M_{\nu QG1} > 2.7 (2.5) \times 10^{10}$~GeV for subluminal
(superluminal) propagation, respectively, and $M_{\nu QG2} > 4.6
(4.1) \times 10^{4}$~GeV at the 95~\% confidence level. These limits
are already much more stringent than those established using the
MINOS detector. We then assess the improved sensitivity to Lorentz
violation that could be obtained if a galactic supernova at a
distance of 10~kpc is observed using the Super-Kamiokande detector,
estimating sensitivities to $M_{\nu QG1}
> 2 (4) \times 10^{11}$~GeV for subluminal (superluminal)
propagation, respectively, and $M_{\nu QG2} > 2 (4) \times
10^{5}$~GeV. All these results are obtained taking neutrino
oscillation effects into account, and assuming that any Lorentz
violation is flavour-independent~\footnote{This is a strong
condition on any model of Lorentz violation, that is imposed by the
success of conventional neutrino oscillation phenomenology, which
implies that flavour-dependent dispersion effects can be neglected
in the analyses of MINOS and OPERA data. Such effects could in
principle appear in neutrinos from supernovae, but would not affect
the results presented below, which are essentially independent of
oscillation hypotheses.}.

We also discuss the sensitivity to Lorentz violation of the OPERA
experiment at the CNGS neutrino beam from
CERN~\footnote{For previous discussions of searches for Lorentz violation in
neutrino data, see~\cite{LeptonNumber,volkov}.}. We recall that the
CNGS beam cycle provides two fast-extracted proton spills lasting
10.5~$\mu$s each and separated by 50~ms, each containing 2100
bunches with standard deviation 0.25~ns, separated from each other
by the CERN SPS RF bucket structure of 5~ns~\cite{Meddahi:2007zz}.
The OPERA data-acquisition (DAQ) system is organized in such a way
that each subdetector provides its data with a distributed
time-stamp with a granularity of 10~ns. If a time-synchronization
method conceptually similar to that of MINOS between the CERN
neutrino extraction-magnet signal and the OPERA time-stamp were
implemented, the sensitivity would be greater than that of MINOS.
This because, even though the baseline between the source and the
detector are the same and the spill lengths are similar, the
neutrinos in the CNGS beam typically have higher energies than those
in the NuMI beam. Exploiting this feature, on the basis of an
optimized analaysis we estimate that after 5 years of running
sensitivities using OPERA could reach
 $M_{\nu QG1} \sim 7 \times 10^{5}$~GeV
($M_{\nu QG2} \sim 8 \times 10^{3}$~GeV) for subluminal
(superluminal) propagation.

Further improvements in sensitivity would result if one could
exploit the RF bucket structure of the spill. Assuming that the
arrival time of the neutrinos would be correlated with the RF bunch
structure with a timing accuracy of, say, 1~ns, the sensitivity to
Lorentz violation could be improved to $M_{\nu QG1}\sim 5 \times
10^{7}$~GeV ($M_{\nu QG2} \sim 4 \times 10^{4}$~GeV) for  
the linear and quadratic cases, respectively. These results could be
improved significantly if neutrino events occurring in the rock
upstream from OPERA could be included in the analysis. In this case,
the sensitivities would become $M_{\nu QG1}\sim 4 \times 10^{8}$~GeV
and $M_{\nu QG2} \sim 7 \times 10^{5}$~GeV. In the case of quadratic
Lorentz violation, this sensitivity is better than that obtained
from supernova 1987a, and even improves on the sensitivity possible
with a future galactic supernova.

%%%%%%%%%%%%%%%%%%%%%%%%%%%%%%%%%%%%%%%%%%%%%%%%%%%%%%%%%%%%%%%%%%%%%%%%%%%%%%%%%%%%%%%
\section{Limits on Lorentz Violation from Supernovae}
%%%%%%%%%%%%%%%%%%%%%%%%%%%%%%%%%%%%%%%%%%%%%%%%%%%%%%%%%%%%%%%%%%%%%%%%%%%%%%%%%%%%%%%
%
In this Section we discus the supernova mechanism and the ability to
test Lorentz violation via the detection of neutrinos created in
this process. We then analyze the data from the supernova SN1987a,
the first supernova from which neutrinos have been detected, giving
bounds at the $95\%$ C.L.. Then we simulate a possible future
galactic supernova and discuss the potential of the next generation
of neutrino detectors, represented by Super-Kamiokande (SK), to
improve this bound.

\subsection{Review of Neutrino Emissions from Supernovae}

The detection of neutrinos from SN1987a in the Large Magellanic
Cloud (LMC) remains a landmark in neutrino physics and astrophysics.
Although only a handful of neutrinos were detected by the
Kamiokande-II (KII)~\cite{k2sn1987a}, Irvine-Michigan-Brookhaven
(IMB)~\cite{imbsn1987a} and Baksan~\cite{baksan1987a} detectors,
they provided direct evidence of the mechanism by which a star
collapses, and the role played by neutrinos in this mechanism
\cite{Totsuka:1991dm}. The numbers and energies of the neutrinos
observed were consistent with the expected supernova energy release
of a few times $10^{53}$~ergs via neutrinos with typical energies of
tens of MeV. A future galactic supernova is expected to generate up
to tens of thousands of events in a water-{\v C}erenkov detector
such as SK, which will clarify further theories of the supernova
mechanism and of particle physics~\cite{Ikeda:2007sa}.

Current simulations reveal several distinct stages of neutrino
emission \cite{Keil:2002in,snnew, Totani:1997vj}. During the early
stage with a typical timescale of a few milliseconds, huge numbers
of $\nu_{e}$ are produced via $pe\rightarrow n\nu_{e}$, known as the
neutronization peak.  Despite the huge numbers of neutrinos
produced, these are difficult to detect water-{\v C}erenkov
detectors, because the neutrinos produced in this process are
detected via scattering on electrons and (in the case of the
electron neutrino) via interactions with Oxygen nuclei. At the
energies of interest the cross section for detection is dominated by
the charged-current interaction $\bar{{\nu}}_{e}p\rightarrow n
e^{+}$, which detects anti-electron neutrinos. During the later
stages of the supernova explosion, all flavours of neutrinos and
antineutrinos are produced with approximate Fermi-Dirac spectra,
that are characterized by different average energies for different
neutrino species: $\langle E_{\nu_{e}}\rangle=(10-12)$~MeV, $\langle
E_{\bar{\nu}_{e}}\rangle=(12-18)$~MeV and $\langle
E_{\nu_{x}}\rangle=(15-28)$~MeV (where $\nu_{x}$ denotes
$\nu_{\mu}$, $\nu_{\tau}$ and their respective antiparticles), with
total emitted energy fractions $\varepsilon_{\nu_{e}}=(10-30)\%$,
$\varepsilon_{\bar{\nu}_{e}}=(10-30)\%$,
$\varepsilon_{\nu_{x}}=(10-20)\%$~\cite{Keil:2002in,snnew}.

The neutrinos produced in the supernova pass from densities close to
nuclear density in the core through to the approximate vacuum of
interstellar space, and the interactions with this matter dominate
the neutrino oscillations. The neutrinos become maximally mixed at
Mikheev-Smirnov-Wolfenstein (MSW) resonances and to first
approximation the nature of the oscillations can be determined by
the properties of these resonances. The resonance condition is
$A=\Delta m^{2}\cos2\theta$, where $A$ is the matter potential,
$\Delta m^{2}$ is the difference in mass squared and $\theta$ is the
mixing angle. For a typical density profile and composition of the
supernova medium, and typical neutrino energies ,the matter
potential is positive (negative) for neutrinos (antineutrinos).
Assuming just the three Standard Model neutrinos, there are two
possible MSW resonances, corresponding to the solar and atmospheric
mass-squared splittings~\cite{msw1,msw2,msw3}. We know from the
solar and KamLAND data that $\Delta m_{21}^{2}\equiv
m_{2}^{2}-m_1^{2}$ is positive, and therefore the corresponding MSW
resonance is in the neutrino sector\cite{solglobal}.

However, the sign of $\Delta m^{2}_{32}$ is undetermined and
therefore the corresponding resonance could be in either the
neutrino or the antineutrino sector, corresponding to the two
possible mass hierarchies, the normal (inverted) for a positive
(negative) $\Delta m^{2}_{32}$. At the resonance there is a
probability of transitions between the mass eigenstates, known as
`level crossing'. If the width of the resonance is large compared to
the neutrino oscillation length at the resonance then the level
crossing probability is small and the resonance is adiabatic. On the
other hand, if the width of the resonance is small compared to the
neutrino oscillation length scale, then transitions between the mass
eigenstates occur and the resonance is said to be non-adiabatic.
Combining current simulations of the supernova and the value of the
solar mixing angle we can determine that the the solar resonance is
adiabatic \cite{Strumia:2006db}. However, the current limit on
$\theta_{13}$ is insufficient to determine whether the atmospheric
resonance is adiabatic or not: simulations indicate that if
$\sin^{2}2\theta_{13}\gtrsim10^{-3}$ the resonance is adiabatic and
if $\sin^{2}2\theta_{13}\lesssim10^{-5}$ the resonance is
non-adiabatic. The oscillation probabilities for both hierarchies
are given in Table \ref{tab:probs}.

In addition to these effects, recent work has shown that neutrino
self-interactions can induce large, non-MSW flavour
oscillations~\cite{Spectral_split}. These occur at large neutrino
densities, just outside the neutrinosphere. For the normal hierarchy
these effects have little effect on the flavour oscillations, but
for the inverted hierarchy with non-zero $\theta_{13}$ significant
flavour changes can occur. These effects result in a `spectral
split', in which the $\nu_e$ and $\nu_x$ spectra are simply swapped
above a critical energy, while the entire spectra of the ${\bar
\nu}_e$ and ${\bar \nu}_x$ are swapped. For the case where the
flavour transformations have occurred before the MSW resonances the
flavour transformations can be thought of as changing the initial
spectra, whereas in the case of shallow density profiles this
becomes more complicated.

We note in addition that, as the shock wave inside the supernova
passes through the atmospheric resonance, it can change it from
adiabatic to non-adiabatic, resulting in a time dependence in the
signal that we do not consider in this
paper~\cite{raffeltforplusrev}.

\begin{table}
\begin{center}
\begin{tabular}{|c|c|c|c|c|c|}
\hline
Hierarchy &  $\sin^{2}\theta_{13}$ & \multicolumn{2}{|c|}{p} & $\bar{p}$ \\
\cline{3-4}
 & & $E<E_{c}$ & $E<E_{c}$ & \\
\hline
Normal             & $\gtrsim 10^{-3}$  & 0                        & 0 & $\cos^{2}\theta_{\odot}$ \\
Inverted           & $\gtrsim 10^{-3}$  & $\sin^{2}\theta_{\odot}$ & $\sin^{2}\theta_{\odot}$ & 0 \\
Normal or Inverted & $\lesssim 10^{-5}$ & $\sin^{2}\theta_{\odot}$ & $\sin^{2}\theta_{\odot}$ & $\cos^{2}\theta_{\odot}$ \\
Inverted SS        & $\gtrsim 10^{-3}$  & $\sin^{2}\theta_{\odot}$ & $\cos^{2}\theta_{\odot}$ & 1 \\
Inverted SS        & $\lesssim 10^{-5}$ & $\sin^{2}\theta_{\odot}$ & $\cos^{2}\theta_{\odot}$ & $\sin^{2}\theta_{\odot}$ \\
\hline
\end{tabular}
\end{center}
\caption{\it The oscillation probabilities for the normal and
inverse hierarchies, including the effect of the spectral split
(SS), where the resulting $\nu_{e}$ and $\bar{\nu}_{e}$ fluxes are
$F_{\nu_{e}}=pF_{\nu_{e}}^{0}+(1-p)F_{\nu_{x}}^{0}$ and
$F_{\bar{\nu}_{e}}=\bar{p}F_{\bar{\nu}_{e}}^{0}+(1-\bar{p})F_{\bar{\nu}_{x}}^{0}$
respectively.}\label{tab:probs}
\end{table}

%%%%%%%%%%%%%%%%%%%%%%%%%%%%%%%%%%%%%%%%%%%%%%%%%%%%%%%%%%%%%%%%%%%%%%%%%%%%%%%%%%%%%%%
\subsection{Analysis Techniques}

%%%%%%%%%%%%%%%%%%%%%%%%%%%%%%%%%%%%%%%%%%%%%%%%%%%%%%%%%%%%%%%%%%%%%%%%%%%%%%%%%%%%%%%

As previously discussed, it has been suggested that QG effects may
lead to Lorentz-violating modifications in the propagation of
energetic particles, and hence to dispersive effects, specifically a
non-trivial refractive index. These dispersive properties of the
vacuum would lead to an energy dependence in the arrival times of
neutrinos. 

{ Even in the absence of any detailed, analytic understanding of time
structure
of a neutrino signal from a supernova, one can exploit the observation that, since the 
neutrino events have a range of energies, an energy dependence of the neutrino
velocity would spread out the arrival times, compared to the signal if
there were no dispersive properties of the vacuum. Any data set comprising 
both the time and energy of each neutrino event can be
analyzed by inverting the dispersion that would be caused by any
hypothesized QG effect. The preferred value of the energy-dependence parameter
would minimize the duration (time spread) of the supernova neutrino signal.} 

Assuming either a linear or a quadratic
form of Lorentz violation: $v/c = [1 \pm (E/M_{\nu QG1})]$ or $[1
\pm (E/M_{\nu QG2})^2]$, a lower limit on $M_{\nu QG1}$ and $M_{\nu
QG2}$ may be obtained by requiring that the emission peak not be
broadened significantly. A non-zero value of $M_{\nu QG1}^{-1}$ or
$M_{\nu QG2}^{-1}$ might be indicated if it reduced significantly
the duration (time spread) of the neutrino signal. The duration
(time spread) of the neutrino signal can be quantified using different
estimators depending mostly on the amount of available statistics and time
profile of the data set, if applicable~\footnote{ Statistically poor event
lists,
such as that for SN1987a, the only one currently available in supernova neutrino 
astronomy, do not allow the time profile to be classified, because time binning is
impractical and one cannot apply nonparametric statistical tests to
unbinned data.}. In the following, we outline two estimators for analyzing
neutrino signals, that we use first to
quantify the limits obtainable from the SN1987a neutrino data and then
the sensitivities that would be provided by a possible future
galactic supernova signal.

\subsubsection{Minimal Dispersion (MD) Method}

{ We assume that the data set consists of a list of neutrino events
with measured energies $E$ and arrival times $t$ such as that in Table~2. 
In the first method, we consider event lists
with a relatively low number of events, that do not allow a
reasonable time profile to be extracted. In this case} we consider
the time
dispersion of the data set, quantified by
\begin{equation}
\label{MD_base}
\sigma_{t}^{2}\equiv\langle\left(t-\langle
t\rangle\right)^{2}\rangle,
\end{equation}
where $t$ is the time of each detected event. We then apply an
energy-dependent time shift $\Delta t=\tau_{l} E^{l}$, where
$\tau_{l}=L/c M_{\nu QGl}^{l}$, varying $M_{\nu QGl}$ so as remove
any assumed dispersive effects. 

{ The `correct' value of the time shift $\tau_{l}$
should always compress the arrival times of the neutrino
events. Any other (`uncorrect') value of $\tau_{l}$ would spread in time the
events relative to the `correct' shift. Therefore, the dispersion~(\ref{MD_base}) can be considered
as an estimator to measure the degree of `compression' of the neutrino events in
time. In the following, we first apply this MD method in a warm-up exercise to
the data from SN 1987a, and later we exhibit in subsection~\ref{future_g_SN}
the typical behaviour of this estimator versus 
$\tau_l$ for hypothetical data from a possible future galactic supernova.

Evaluating the dispersion~(\ref{MD_base}) one obtains}
\begin{eqnarray}
\label{disp_tau}
\sigma_{t}^{2}&=&\langle \left(t - \tau_{l} E^{l} - \langle t -
\tau_{l} E^{l}\rangle\right)^{2}\rangle \\ & = & \langle t^{2}
\rangle- \langle t \rangle^{2} - 2 \tau_{l} \left(\langle t
E^{l}\rangle - \langle t\rangle \langle E^{l}\rangle
\right)+\tau_{l}^{2}\left(\langle E^{2l}\rangle-\langle
E^{l}\rangle^{2}\right).
\end{eqnarray}
Therefore, the dispersion of the `de-refracted' time distribution
is minimized by the parameter
$\tau_{l}^{min}$, defined by
\begin{equation}
\tau_{l}^{min} \equiv \frac{\langle t E^{l}\rangle- \langle t\rangle
\langle E^{l}\rangle}{\langle E^{2l}\rangle-\langle
E^{l}\rangle^{2}} . \label{eq:taumin}
\end{equation}
We can use (\ref{eq:taumin}) for any data set to estimate the scale
$M_{\nu QGl}$ at which Lorentz violation is manifest. However, there
are uncertainties in the energy and time measurements, as well as
statistical uncertainties in the estimation of the observables
calculated from any given data set, compared to their true values.
We estimate the statistical uncertainties of an observable $x$ as
\begin{equation}
\sigma_{x}^{stat}=\sqrt{\frac{\langle x^{2}\rangle-\langle
x\rangle^{2}}{N}}, \label{eq:tauuncert}
\end{equation}
where N is the number of events, and $x=E, t$ or some combination of
both. In order to estimate the uncertainties in $\tau_{l}^{min}$, we
use a Monte Carlo simulation to repeat the calculation of
$\tau_{l}^{min}$ including the energy and statistical uncertainties.
We then make a Gaussian fit and use it to quote best-fit parameters
and errors.
%
%%%%%%%%%%%%%%%%%%%%%%%%%%%%%%%%%%%%%%%%%%%%%%%%%%%%%%%%%%%%%%%%%%%%%%%%%%%%%%%%%%%%%%%
\subsubsection{Energy Cost Function (ECF) Method}
%%%%%%%%%%%%%%%%%%%%%%%%%%%%%%%%%%%%%%%%%%%%%%%%%%%%%%%%%%%%%%%%%%%%%%%%%%%%%%%%%%%%%%%

{ This is a different analysis technique that is mostly applicable to event
lists
that are statistically rich. This means
that one can combine the neutrino events into a time profile exhibiting
pulse features that can be distinguished (parametrically or nonparametrically)
from a uniform distribution at high confidence level.} 

For the analysis we first
choose the most
active (transient) part of the signal $(t_{1};t_{2})$, as defined using a
Kolmogorov-Smirnov (KS) statistic. The KS statistic is calculated
using the difference between the cumulative distribution function
(CDF) of the unbinned data and that of a uniform distribution. { The
KS statistic is defined as the time that elapses between the minimum
and maximum of this difference~\footnote{ The most active part of the signal
can be also chosen by fitting the binned time profle, but the
nonparametric way we use to extract a feature is less dependent on the time profile.
In the case of a multipulse structure  of the time profile, several windows may
be analized separately.}.} { Having chosen this window, we scan over its whole
support the time distribution
of all events, shifted by $\Delta t=\tau_{l} E^{l}$, and sum the energies 
of events in the
window. This procedure is repeated for many values of $\tau_l$,
chosen so that the shifts $\Delta t$ match the precision of the arrival-time
measurements, thus defining the `energy cost function' (ECF).
The maximum of the ECF indicates the value of $\tau_l$ that best  recovers
the signal, in the sense of maximizing its power (amount of energy in a
window of a given time width $t_{2}-t_{1}$). This
procedure is then repeated for many Monte-Carlo (MC)
data samples generated by
applying to the measured neutrino energies the estimated Gaussian
errors. A typical ECF for one particular MC realization as well as the
typical distribution of the positions of the maxima of the ECFs for many enegy
MC realizations are illustrated in subsection~\ref{future_g_SN}
(see Fig.~\ref{ECF} and Fig.~\ref{ECF_position}). }
 
We perform this procedure for
different energy weightings $E^{n}$, where n=0,1,2, { summing up either
the numbers of events, the energies or the squares of the energies in
the time window selected}, so to optimize the
errors placed on the scale of Lorentz violation. 

%
%%%%%%%%%%%%%%%%%%%%%%%%%%%%%%%%%%%%%%%%%%%%%%%%%%%%%%%%%%%%%%%%%%%%%%%%%%%%%%%%%%%%%%%
\subsection{Data Analysis}

For the analysis of SN1987a we use the uncertainties in Table~2,
which were taken from~\cite{Loredo:2001rx}. In the case of a
possible galactic supernova, we consider the Super-Kamiokande (SK)
water Cerenkov detector, and we use the detector properties given
in~\cite{raffeltforplusrev,Tomas:2003xn}, where the energy
uncertainties are modelled as $\sigma_{E}^{det 2}=\sqrt{E_{0}E}$,
where $E_{0}=0.22$~MeV. We note that the uncertainties in the time
measurements are in general much less than the statistical and
energy uncertainties, and we therefore neglect them in our analysis.

\begin{table}[h]
\begin{center}
\begin{tabular}{c c}
\begin{tabular}{|c|c|c|}
  \hline
  \multicolumn{3}{|c|}{IMB}\\
  \hline
  t (s) & E (MeV) & $\sigma_{E}$ (MeV) \\
  \hline
  $t\equiv0.0$ & 38 & 7 \\
  0.412 & 37 & 7 \\
  0.650 & 28 & 6 \\
  1.141 & 39 & 7 \\
  1.562 & 36 & 9 \\
  2.684 & 36 & 6 \\
  5.010 & 19 & 5 \\
  5.582 & 22 & 5 \\
  \hline
  \multicolumn{3}{|c|}{Baksan}\\
  \hline
  t (s) & E (MeV) & $\sigma_{E}$ (MeV) \\
  \hline
  $t\equiv0.0$   & 12.0 & 2.4 \\
  0.435 & 17.9 & 3.6 \\
  1.710 & 23.5 & 4.7 \\
  7.687 & 17.6 & 3.5 \\
  9.099 & 10.3 & 4.1 \\
  \hline
\end{tabular}
&
\begin{tabular}{|c|c|c|}
  \hline
  \multicolumn{3}{|c|}{Kamiokande II}\\
  \hline
  t (s) & E (MeV) & $\sigma_{E}$ (MeV) \\
  \hline
  $t\equiv0.0$   & 20.0 & 2.9 \\
  0.107 & 13.5 & 3.2 \\
  0.303 & 7.5  & 2.0 \\
  0.324 & 9.2  & 2.7 \\
  0.507 & 12.8 & 2.9 \\
  1.541 & 35.4 & 8.0 \\
  1.728 & 21.0 & 4.2 \\
  1.915 & 19.8 & 3.2 \\
  9.219 & 8.6  & 2.7 \\
  10.433 & 13.0& 2.6 \\
  12.439 & 8.9 & 1.9 \\
  \hline
  \multicolumn{3}{c}{}\\
  \multicolumn{3}{c}{}\\
  \multicolumn{3}{c}{}\\
  \multicolumn{3}{c}{}\\
\end{tabular}
\\
\end{tabular}
\end{center}\label{tab:data}
\caption{\it The measured neutrino data from SN1987a, where we have
omitted the events identified previously as background, and in each
data set we define $t\equiv0.0s$ for the first event.}
\end{table}

\subsubsection{SN1987a}\label{subsection:sn1987}
\begin{figure}[h]
\begin{center}
\includegraphics[width=0.8\textwidth]{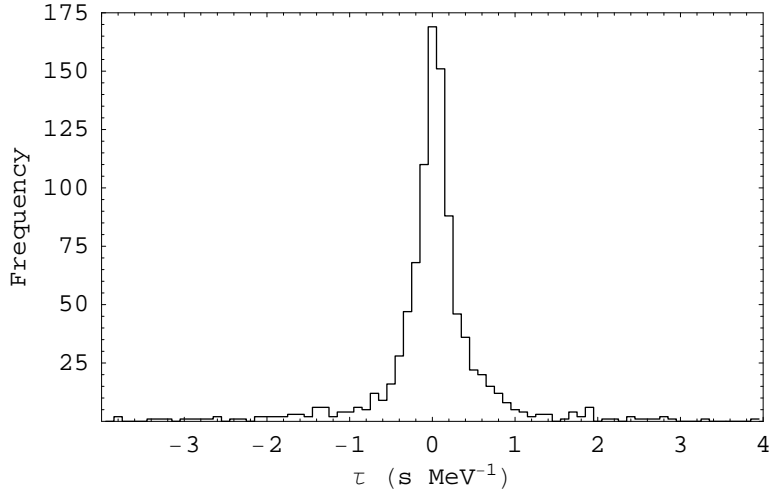}
\end{center}\label{fig:KII}
\caption{\it The distribution $\tau_{min}$ of 1000 Monte Carlo
simulations of the KII data on neutrinos from SN1987a, including the
smearing due to energy uncertainties.} \label{fig:SN1987aKII}
\end{figure}

%%%%%%%%%%%%%%%%%%%%%%%%%%%%%%%%%%%%%%%%%%%%%%%%%%%%%%%%%%%%%%%%%%%%%%%%%%%%%%%%%%%%%%%
%
Neutrinos from SN1987a were detected in three detectors, KII, IMB
and Baksan. The times and energies of the events are given in
Table~2. The minimum dispersion was calculated 1000 times for each
data set to include the smearing from uncertainties. As an example,
Fig.~\ref{fig:SN1987aKII} shows this smearing for the KII data set.
From these distributions we can determine the best fit and the
error, which are summarized in Table \ref{tab:SN1987a}. We analyze
similarly the data from IMB and Baksan. As there is an uncertainty
in the relative time measurements of each detector, we analyze each
data set independently using the minimal dispersion method and then
combine them to quote the final best fit and error, as shown in
Table~\ref{tab:SN1987a}.

\begin{table}[h]
\begin{center}
\begin{tabular}{|c|c|c|c|c|}
   \hline
   Data set & \multicolumn{2}{|c|}{$\tau_{1}({\rm s\cdot MeV^{-1})}$} &
\multicolumn{2}{|c|}{$\tau_{2}(10^{-3}{\rm s\cdot MeV^{-2}})$} \\
   \cline{2-5}
    & Best fit & Error & Best fit & Error \\
   \hline
   Kamiokande II & -0.0233307  & 0.197601 & -0.685 & 2.935 \\
   IMB           & -0.00417622 & 0.121513 & -0.308 & 1.601 \\
   Baksan        &  0.0574167  & 0.47789  &  2.704 & 8.105 \\
   \hline
   Combined      & -0.00643648 & 0.101162 & -0.304 & 1.385 \\
   \hline
 \end{tabular}
 \end{center}
 \caption{\it The best fits to the SN1987a neutrino data obtained using the
 minimal dispersion method.}\label{tab:SN1987a}
\end{table}

\begin{figure}[h]
\centering
\begin{tabular}{cc}
\epsfig{file=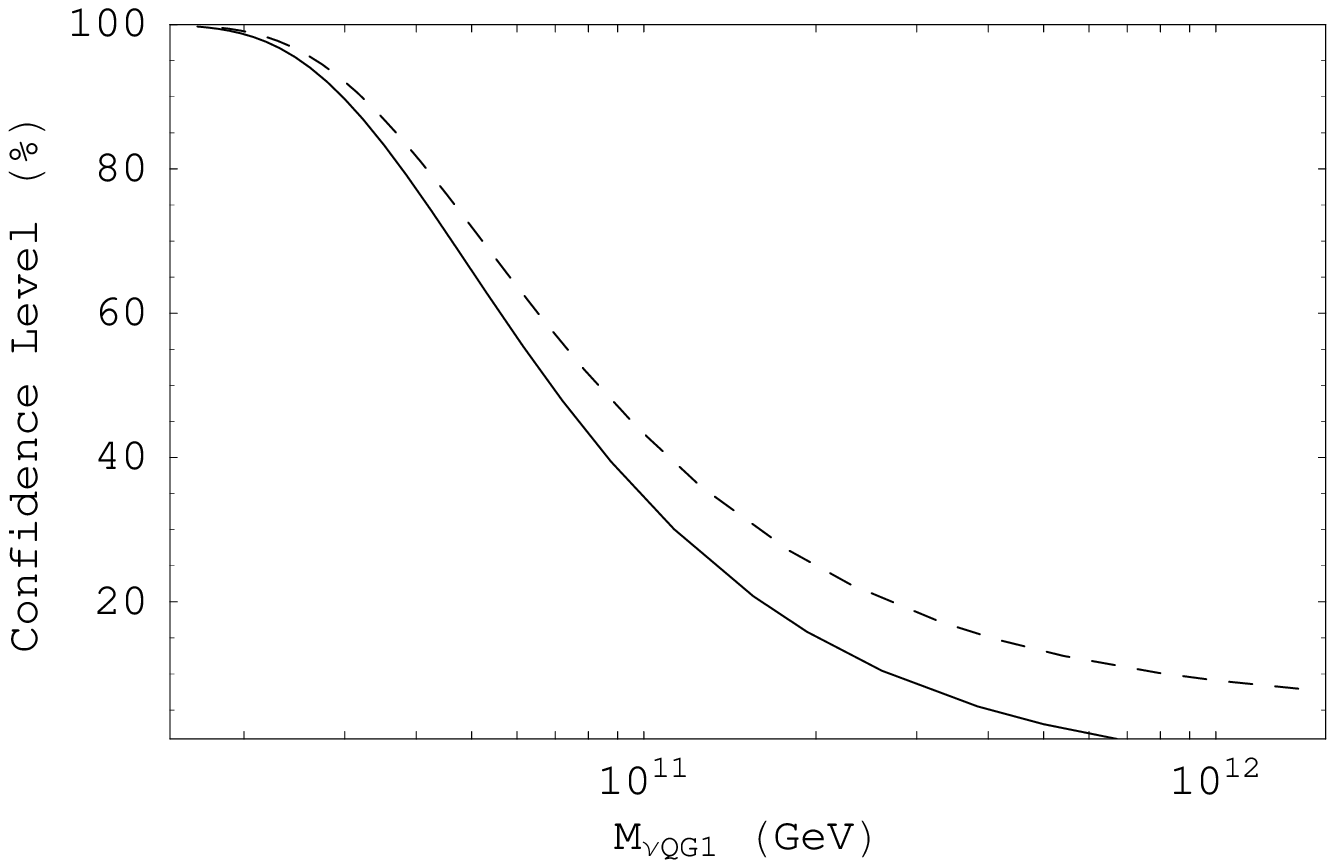,width=0.4\linewidth} &
\epsfig{file=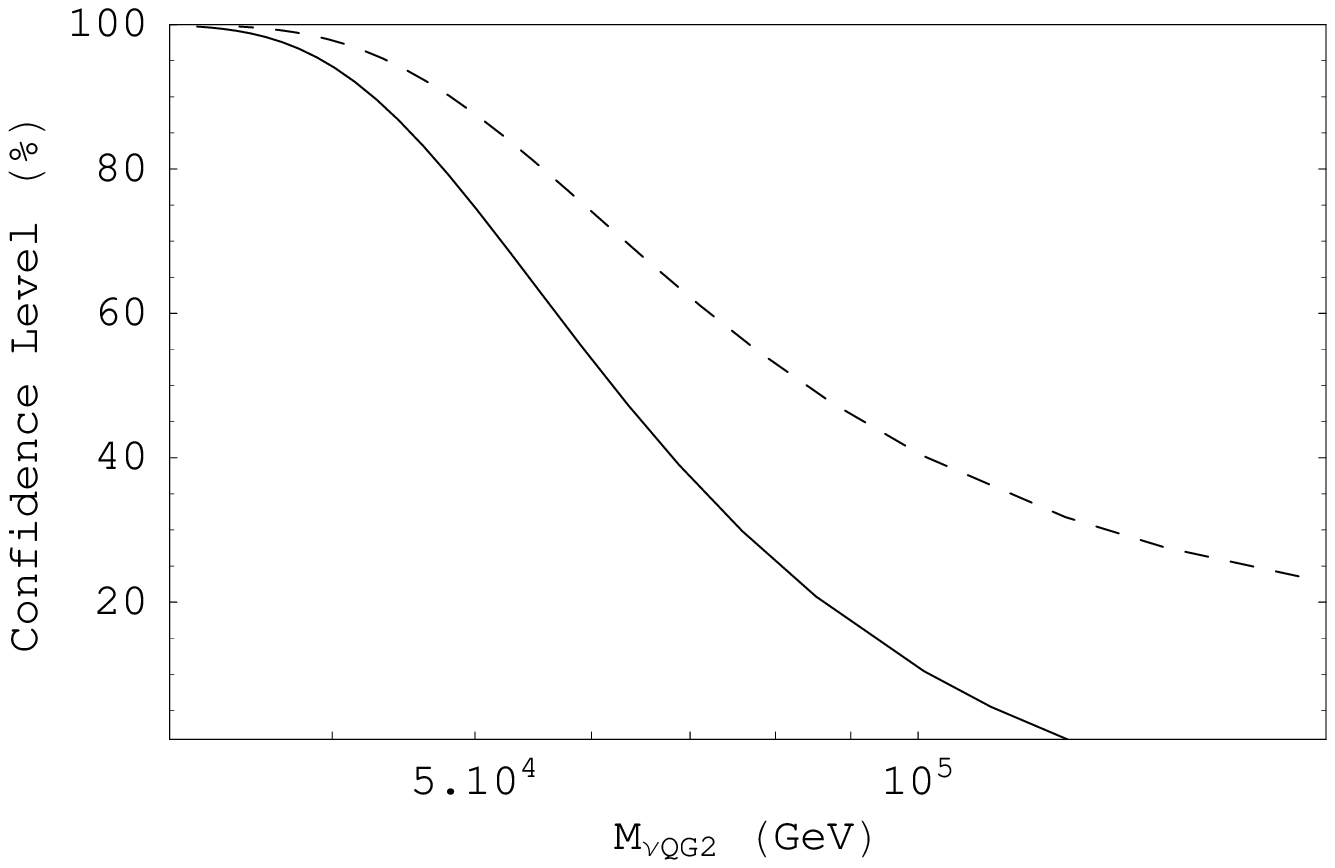,width=0.4\linewidth} \\
\end{tabular}
\caption{\it The regions of parameter space excluded by SN1987a, for
subluminal (dashing) and superluminal (black) linear (left) and
quadratic (right) models.} \label{fig:SN1987aCL}
\end{figure}

On the basis of this combined analysis, Fig.~\ref{fig:SN1987aCL}
shows the region which is excluded by the SN1987a data. Taking the
distance to the supernova as $L=(51.3\pm1.2)$~kpc, the scale at
which Lorentz violation may enter the neutrino sector is constrained
to be
\begin{equation}
M_{\nu QG1} > 2.7 \times 10^{10}~{\rm GeV~or}~M_{\nu QG1}
> 2.5 \times 10^{10}~{\rm GeV}
\end{equation}
at the $95\%$ C.L. for the linear subluminal and superluminal models
respectively. The corresponding limits for the quadratic models are
\begin{equation}
M_{\nu QG2} > 4.6 \times 10^{4}~{\rm GeV~or}~M_{\nu QG2} > 4.1
\times 10^{4}~{\rm GeV}
\end{equation}
at the $95\%$ C.L. for the subluminal and superluminal versions,
respectively.

%%%%%%%%%%%%%%%%%%%%%%%%%%%%%%%%%%%%%%%%%%%%%%%%%%%%%%%%%%%%%%%%%%%%%%%%%%%%%%%%%%%%%%%
\subsection{A Possible Future Galactic Supernova} \label{future_g_SN}
%%%%%%%%%%%%%%%%%%%%%%%%%%%%%%%%%%%%%%%%%%%%%%%%%%%%%%%%%%%%%%%%%%%%%%%%%%%%%%%%%%%%%%%

\begin{figure}[h]
\centering \epsfig{file=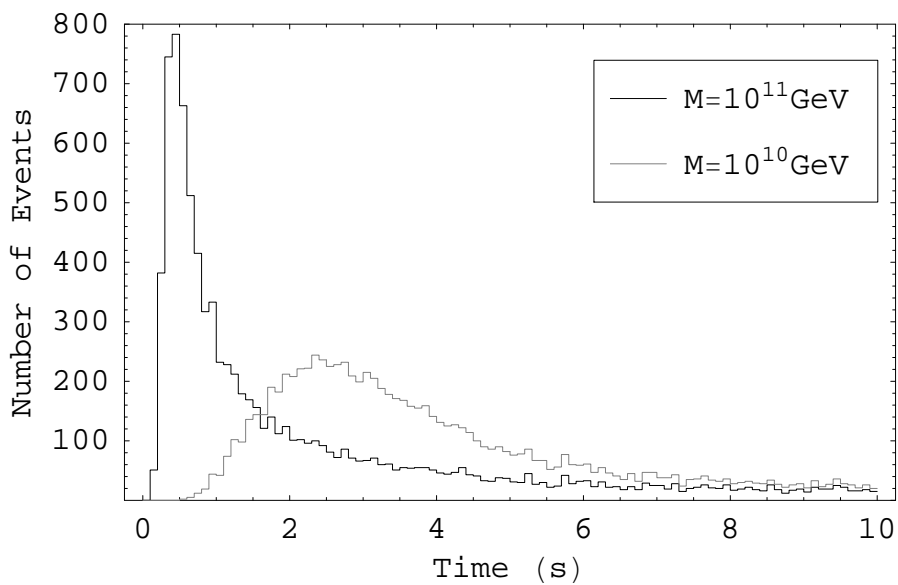,width=0.8\linewidth}
\caption{\it The time distribution of events predicted by our Monte
Carlo simulation for the case of subluminal Lorentz violation at the
mass scales $M=10^{10}GeV$ and $M=10^{11}GeV$.} \label{fig:events}
\end{figure}

\begin{figure} [htb]
\begin{center}
\epsfig{file=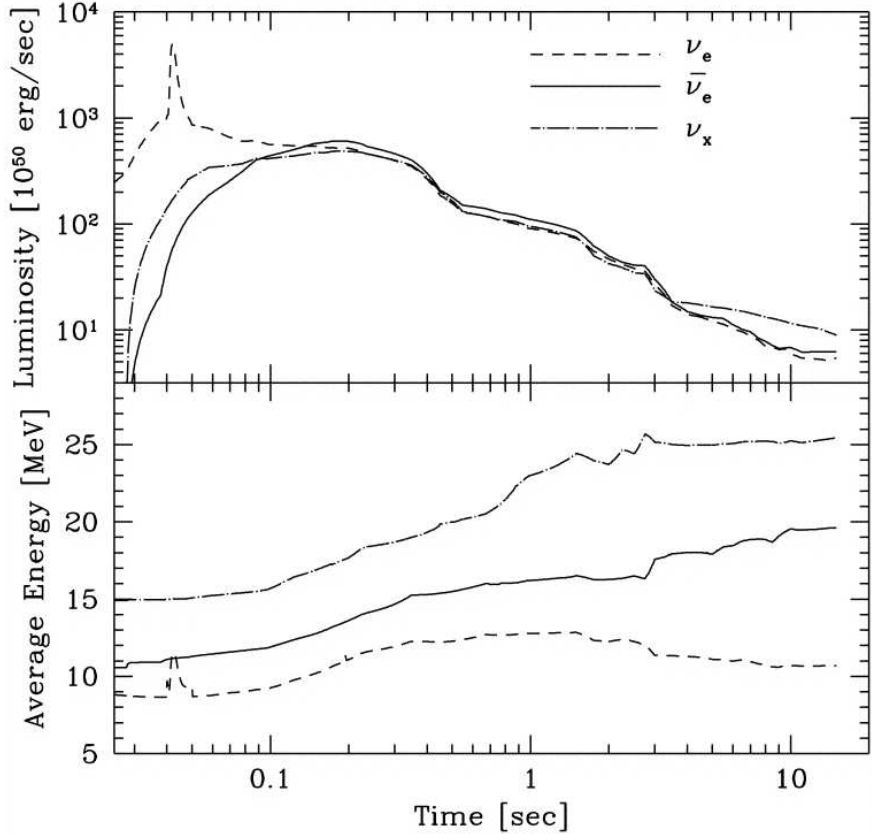,width=0.7\textwidth}
\end{center}
\caption{\it The neutrino energy spectra from the Livermore
simulation~\cite{Totani:1997vj}.} \label{fig:SNSpectrum}
\end{figure}

The detection of a galactic supernova would provide improved
sensitivity to the scale at which Lorentz violation might enter the
neutrino sector, due to an increase in the number of neutrinos which
would be detected. The number of events would also increase because
the current neutrino detectors are larger than those used to detect
neutrinos from SN1987a. However, these effects would be partially
offset because $\tau_{l}\propto L$ and therefore the time-energy
shift will be reduced if, as expected, the supernova takes place
within the galactic disc at a distance $\sim 10$~kpc, compared to
SN1987a in the LMC at a distance of $\sim 51$~kpc. The increase in
the number of neutrinos which are expected to be detected are also
because the next supernova is expected to be closer to the earth
than SN1987a. For definiteness, we use here a Monte Carlo simulation
of the Super-Kamiokande (SK) neutrino detector, but note that other
neutrino detectors could also probe this
physics~\cite{NeutrinoDetectors}. Simulations estimate that the
number of events detected in SK from a supernova at 10~kpc would be
of the order of 10,000 \cite{Ikeda:2007sa}. In order to analyze at
what scales Lorentz violation could be probed by the detection of
galactic supernova neutrinos, we made Monte Carlo simulations with
various levels of linear and quadratic Lorentz violation. We used
the energy spectra of neutrinos from the Livermore
simulation~\cite{Totani:1997vj}, which is shown in Figure
\ref{fig:SNSpectrum} and the detector properties given
in~\cite{Tomas:2003xn}.

We show in Fig.~\ref{fig:events} results from our Monte Carlo
simulation including both charged-current and neutral-current events
for linear sublminal Lorentz violation at the energy scales $M_{\nu
QG1}=10^{10}GeV$ and $M_{\nu QG1}=10^{11}GeV$, including
oscillations corresponding to the normal hierarchy and assuming that
the atmospheric resonance is adiabatic. The signal has spread out
and shifted in time, as we would expect. This time shift is
unobservable because it is shifted relative to the signal in the
absence of Lorentz violation, which in practice cannot be measured.
We have applied the MD and the maximal ECF methods with various
energy weightings to the Monte Carlo data with $M_{QG1}=10^{10}$~GeV
in order to estimate the level of Lorentz violation. 

\begin{figure} [htb]
\begin{center}
\epsfig{file=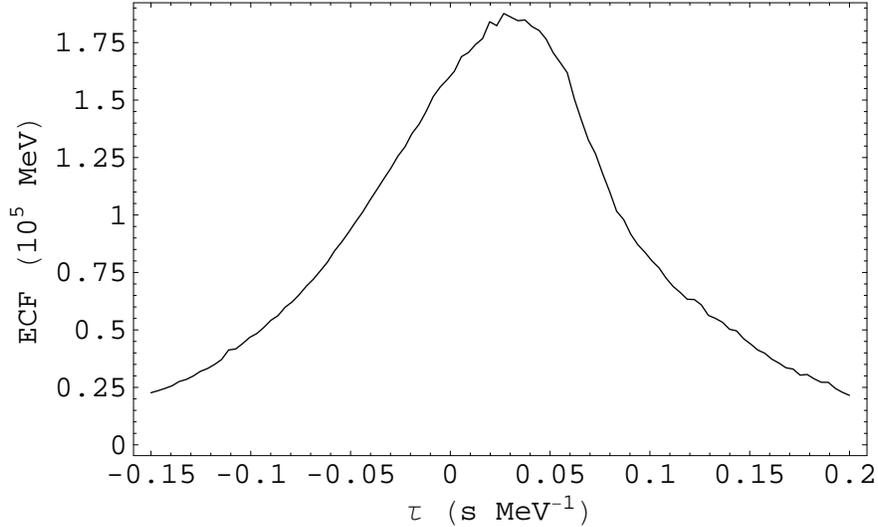,width=0.7\textwidth}
\end{center}
\caption{\it The ECF linearly weighted with energy from one realization of the
simmulated time profile Fig.~\ref{fig:events} with neutrino energies smeared by
MC applying to the expecteed energy resolution of SK, for the case of
linear energy depending neutrino velocity.} \label{ECF}
\end{figure}

\begin{figure} [htb]
\begin{center}
\epsfig{file=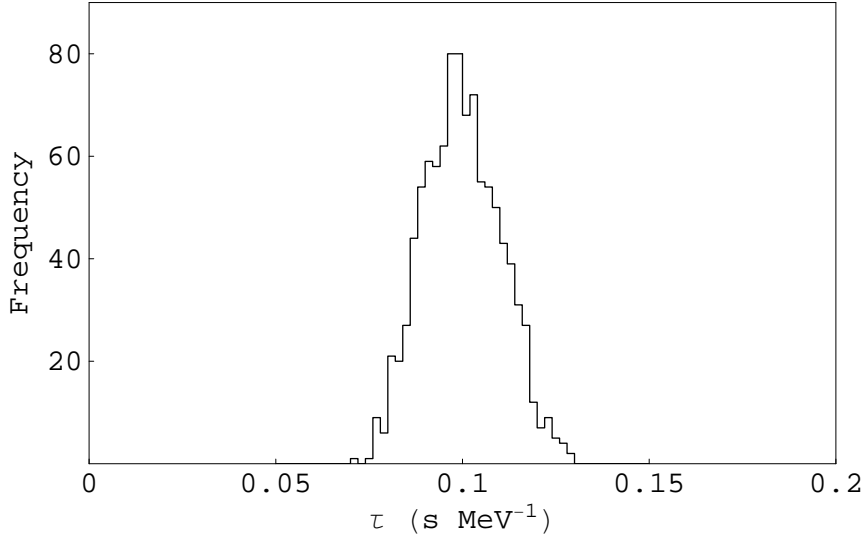,width=0.7\textwidth}
\end{center}
\caption{\it The distribution of the postions of the maximums from fits of ECFs
like in~Fig.~\ref{ECF} of 1000 realizations of the simmulated time profile
Fig.~\ref{fig:events} with neutrino energy smeared by MC.} \label{ECF_position}
\end{figure}

Fig.~\ref{ECF} shows the
ECF for one realization of the energy-smeared sample obtained applying MC to the
measured neutrino energies with the Gaussian errors expected from SK.
It exhibits a clear maximum, whose position may be estimated by fitting it
with a Gaussian profile in the peak vicinity. Fig.~\ref{ECF_position} shows the 
results of such fits to the
ECFs constructed for the 1000
energy-smeared realizations. From this distribution we can
 derive
the prefered value of
$\tau_l$.

\begin{figure} [htb]
\begin{center}
\epsfig{file=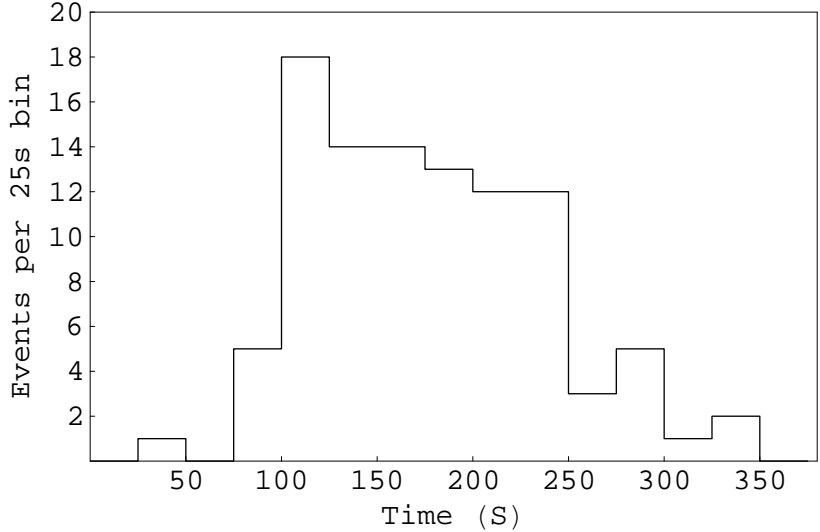,width=0.7\textwidth}
\end{center}
\caption{\it The time profile of 600 events, which could be detected by SK from
a
future extra galactic supernovae occured at the distance 40~kpc from the earth.
The events are simmulated with respect to the energy spectrum given
in~Fig.~\ref{fig:SNSpectrum} with linear energy depending propagation
effect, encoded at the level $\tau_1 = 5.5\ s\cdot MeV^{-1}$. } \label{600}
\end{figure}

\begin{figure} [htb]
\begin{center}
\epsfig{file=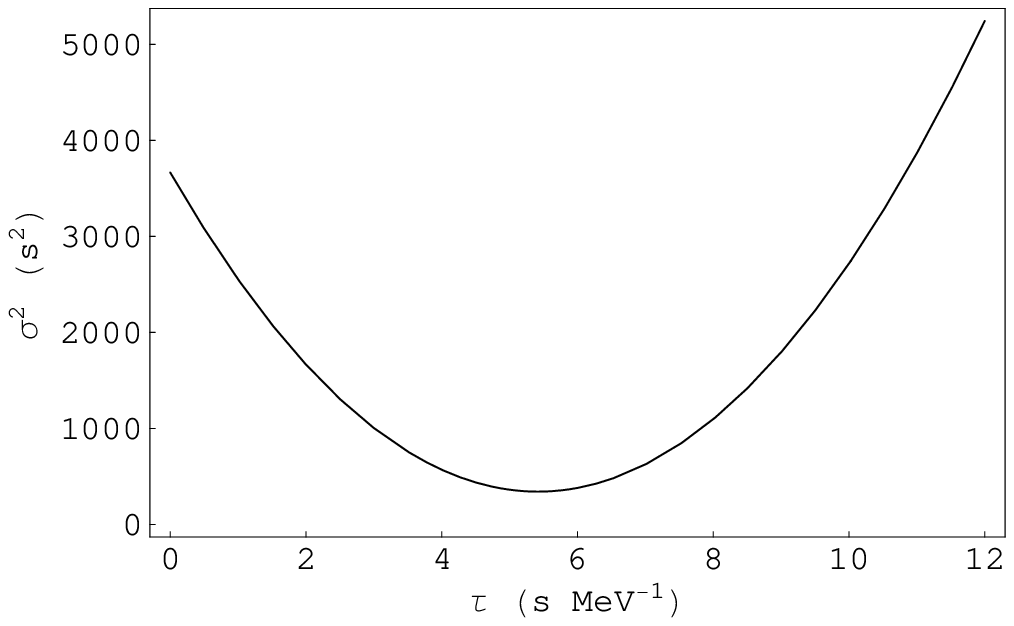,width=0.7\textwidth}
\end{center}
\caption{\it The dispersion (\ref{disp_tau}) versus $\tau$ from
one realization of the simmulated time profile Fig.~\ref{600} with
neutrino energies smeared by
MC applying to the expecteed energy resolution of SK, for the case of
linear energy depending neutrino velocity.}
\label{disp600}
\end{figure}

The results are
sumarized in Table~\ref{tab:MCmethod}, where we have defined ${\hat
m_{l}} \equiv M_{\nu QGl}/M_{\nu QGl}^{true}$, where $M_{\nu
QGl}^{true}$ is the true scale of Lorentz violation and $M_{\nu
QGl}$ is that deduced from the analysis method. Comparing these
results, we find that the maximal ECF technique has greater
sensitivity than the MD method, and that the linear energy weighting
has the greatest sensitivity among the ECF analyses. We therefore
use this in the following.

We have performed simulations for both the normal and inverted mass
hierarchies, with and without the spectral splits caused by neutrino
self-interactions, for the extreme cases $P_{H}=0.0$ and
$P_{H}=1.0$, and analysed them using the ECF method. The
corresponding results are summarized in
Table~\ref{tab:MChierarchies}, where we see that Lorentz violation
can be probed with similar sensitivity for all mass hierarchies.

\begin{table}[h]
\begin{center}
\begin{tabular}{|c|c|}
  \hline
  Method & $95\%$ C.L. \\
  \hline
  Minimal Dispersion (MD) & $0.60 < {\hat m_{1}} < 2.37$ \\
 ECF 0th order   & $0.90 < {\hat m_{1}} < 1.29$ \\
 ECF 1st order   & $0.88 < {\hat m_{1}} < 1.26$ \\
 ECF 2nd order   & $0.87 < {\hat m_{1}} < 1.27$ \\
  \hline
\end{tabular}
\caption{\it The $95\%$ C.L. ranges of ${\hat m_{l}} \equiv M_{\nu
QGl}/M_{\nu QGl}^{true}$ obtained using the different dispersion
methods and various energy weights for a Monte Carlo simulation of a
possible future galactic supernova for $P=0.0$, assuming the normal
mass hierarchy and $M_{\nu QG1}=10^{10}$~GeV.} \label{tab:MCmethod}
\end{center}
\end{table}

\begin{table}[h]
\begin{center}
\begin{tabular}{|c|c|}
  \hline
  Mass hierarchy & $95\%$ C.L. \\
  \hline
  NH $P=0.0$ & $ 0.90 < {\hat m_{1}} < 1.29$ \\
  NH $P=1.0$ & $ 0.90 < {\hat m_{1}} < 1.28$ \\
  IH $P=0.0$ & $ 0.91 < {\hat m_{1}} < 1.26$ \\
  IH SS $P=0.0$ & $ 0.90 < {\hat m_{1}} < 1.27$ \\
  IH SS $P=1.0$ & $ 0.91 < {\hat m_{1}} < 1.28$ \\
  \hline
\end{tabular}
\caption{\it The $95\%$ C.L. for ${\hat m_{l}} \equiv M_{\nu
QGl}/M_{\nu QGl}^{true}$ obtained using the ECF method for a Monte
Carlo simulation of a possible future galactic supernova, for the
normal (NH) and inverted hierarchies (IH), and including the effect
of a spectral split (SS), where $P$ is the level-crossing
probability, and NH $P=1.0$ is equivalent to IH $P=1.0$.}
\label{tab:MChierarchies}
\end{center}
\end{table}

The top three rows of Table~\ref{tab:MClinearQuad} show the results
of our analysis for the linear cases $M_{\nu QG1}=(10^{10}, 10^{11},
10^{12})$~GeV, using the minimal ECF method with no energy
weighting, and making linear and quadratic fits. We see that data
from a future galactic supernova could place strong $95\%$ C.L.
limits on the range of $M_{\nu QG1}$ if it is lower than
$10^{11}$~GeV. In the limit of negligible Lorentz violation ($M_{\nu
QG1} \ge 10^{12}$~GeV), we find the lower limits $M_{\nu QG1}> 2.2
\times 10^{11}GeV$ and $M_{\nu QG1}> 4.2 \times 10^{11}GeV$ at the
$95\%$ C.L. for subluminal and superluminal models, respectively.
The bottom three rows of Table~\ref{tab:MClinearQuad} show the
corresponding results for the quadratic cases $M_{\nu
QG2}=(10^{4.5}, 10^{5}, 10^{5.5})$~GeV, again using the minimal ECF
method with no energy weighting. We see that data from a future
galactic supernova could place strong $95\%$ C.L. limits on the
range of $M_{\nu QG2}$ if it is lower than $10^{5}$~GeV. In the case
of large $M_{\nu QG2}$, we find the lower limits $M_{\nu QG2}> 2.3
\times 10^{5}$~GeV and $M_{\nu QG2}> 3.9 \times 10^{5}$~GeV at the
$95\%$ C.L. for subluminal and superluminal models, respectively, in
the quadratic case.

\begin{table}[h]
\begin{center}
\begin{tabular}{|c|c|}
  \hline
  Model & $95\%$ C.L. \\
  \hline
  $M_{\nu QG1}=10^{10}{\rm GeV}$ & $ 0.90 < {\hat m_{1}} < 1.29$ \\
  $M_{\nu QG1}=10^{11}{\rm GeV}$ & $ 0.64 < {\hat m_{1}} < 1.93$ \\
  $M_{\nu QG1}=10^{12}{\rm GeV}$ & $ 0.22 < {\hat m_{1}},~ 0.42 < {\hat m_{1}^{super}}$
\\
  $M_{\nu QG2}=10^{4.5}{\rm GeV}$ & $0.93 < {\hat m_{2}} < 1.23$\\
  $M_{\nu QG2}=10^{5}GeV$   & $ 0.65 > {\hat m_{2}},~ 2.3  < {\hat m_{2}^{super}}$\\
  $M_{\nu QG2}=10^{5.5}GeV$ & $ 0.19 > {\hat m_{2}},~ 0.72 < {\hat m_{2}^{super}}$\\
  \hline
\end{tabular}
\caption{ \it
The $95\%$ C.L. limits on $M_{\nu QG1}$ and $M_{\nu QG2}$ obtained
using the KS statistic and the ECF method, for subluminal Lorentz
violation with certain input choices of $M_{\nu QG1}$ (top three
rows) and $M_{\nu QG2}$ (bottom three rows). We give the $95\%$ C.L.
limits for subluminal (superluminal) propagation as ${\hat m_{1}}$
(${\hat m_{1}^{super}}$); if a limit for ${\hat m_{1}^{super}}$ is
not given then superluminal propagation has been ruled out at the
$95\%$ C.L..} \label{tab:MClinearQuad}
\end{center}
\end{table}

{ Although the ECF
method is more sensitive than the
MD method, it is not applicable to a statistically poor data set.
The ECF method is best for the analysis of a feature
in a distribution superposed on a uniform background, and the
extraction procedure is possible only with a relatively
representative (i.e., large) sample of events. This is demonstrated by
simmulating a possible future extra galactic supernova which might take place
at a distance similar to that of SN 1987a. The simulation has been
performed in such a way as to have a sample with sufficient statistics to claim at least a
3-$\sigma$ detection of Lorentz invariance in
neutrino propagation. This would need about 600 events
for the linear case, corresponding, assuming the sensitivity of SK,
to a supernova at a distance of about 40~kpc from the
Earth. An expected time profile is presented in Fig.~\ref{600}. The
signal Fig.~\ref{600} contains 600 events and the time distribution encodes
a linearly energy-dependent propagation effect at the level of $\tau_1 = 5.5\
s\cdot MeV^{-1}$, corresponding to $M_{\nu QG1}=7 \times 10^{9}\ GeV$. This
distribution does not demonstrate any significant feature that one could extract in a
time window to be analyzed using the ECF. Therefore, we apply the MD method, which
is better for a signal with poor statistics. The typical behaviour of the
dispersion~(\ref{disp_tau}) versus $\tau$ for one realization of the
energy-smeared sample of the 600 simulated events is presented
in Fig.~\ref{disp600}. The distribution of $\tau_{min}$ given by
(\ref{eq:taumin}) of 1000 MC simulations similar to
Fig.~\ref{fig:SN1987aKII} recovers, in this case, the encoded signal
$\tau_1 = 5.5\ s\cdot MeV^{-1}$ ($M_{\nu QG1}=7 \times 10^{9}\ GeV$) at the
3-$\sigma$ level. A similar simulation for
the quadratic case would require about 400 events, which would
correspond to a SN at a distance of about 50~kpc from the Earth for the
SK efficiency. A 3-$\sigma$ signal could be recovered if the dispersion effect
was at the level $\tau_2 = 0.1\ s\cdot MeV^{-2}$, which corresponds to $M_{\nu
QG2}=7 \times 10^{3}\ GeV$.  

The minimal 3-$\sigma$ discovery statistics, which amounts to
600 (400) events for linear (quadratic) energy dependence, is defined for the MD
method by the uncertainty in the denominator of (\ref{eq:taumin}), which reads
$\approx 5/\sqrt{N}$ ($\approx 4/\sqrt{N}$) 
for either the simulated events or events from SN 1987a, where $N$
is the number of detected events. This means, that for the statistics of
SN 1987a, a Lorentz-violating signal could be detected only at about 
the 1-$\sigma$ C.L., corresponding to the bounds obtained in the previous
subsection. In the case of limited statistics like SN 1987a,
It is possible to estimate similar limits on Lorentz violation without the full
MD machinery used
in~\ref{subsection:sn1987}. However, such an estimate would implicitly
assumes that the dispersion of the initial signal at the source is known. One
could rely on computer simulations of a supernova
explosion~\cite{Totani:1997vj}, but this would introduce an element of 
model-dependent information into the analysis. The methods considered here do not
assume any knowladge on the true profile (spread) of the neutrino signal at the
source: instead, they remove any propagation effect that may be encoded in the
time profile.}

%%%%%%%%%%%%%%%%%%%%%%%%%%%%%%%%%%%%%%%%%%%%%%%%%%%%%%%%%%%%%%%%%%%%%%%%%%%%%%%%%%%%%%%
\section{CNGS and the OPERA Experiment}
%%%%%%%%%%%%%%%%%%%%%%%%%%%%%%%%%%%%%%%%%%%%%%%%%%%%%%%%%%%%%%%%%%%%%%%%%%%%%%%%%%%%%%%

In this Section we discuss the sensitivities to Lorentz violation in
neutrino propagation that could be provided by the OPERA experiment
in the CNGS neutrino beam. We first discuss the sensitivity to
Lorentz violation that could be obtained using the spill structure
alone, without taking into account its bunch substructure. In a
second step, we consider how this bunch substructure could be
exploited to improve the sensitivity, which could be possible if the
timing resolution currently expected for the OPERA detector could be
improved significantly.

We first recall some of the details of the pioneering analysis of
the neutrino velocity in a long-baseline neutrino beam that has been
published by the MINOS collaboration using the NuMI
beam~\cite{Rebel:2008th}. This analysis compared the absolute
timings of the detected neutrino events in the near and far
detectors. The arrival times in the near detector provide a direct
measurement of the neutrino intensity time-profile, consisting of
either 5 or 6 batches separated by short gaps within a 9.7~$\mu s$
long spill. The near and far clocks were synchronized absolutely by
means of Global Positioning Satellite (GPS) receivers. The resulting
systematic error of $\pm$64~ns was dominated by uncertainties in the
delays in the optical fibres that ran between the surface antennae
and the underground detectors. Including the jitter of the two GPS
clocks, the total relative time uncertainty was $\sigma=$150~ns.
This analysis measured $(v-c)/c = (5.1 \pm 2.9) \times 10^{-5}$ at
the 68\% C.L., or $-2.4 \times 10^{-5} < (v - c)/c < 12.6 \times
10^{-5}$ at the 99\% C.L., at an average neutrino energy of 3~GeV
\cite{Rebel:2008th}. In the case of linear Lorentz violation, this
would correspond approximately to $M_{\nu QG1} > 1.2 (4.2) \times
10^5$~GeV in the case of subluminal (superluminal) propagation.

\subsection{CNGS Beam Characteristics}

\begin{figure} [htb]
\begin{center}
\epsfig{file=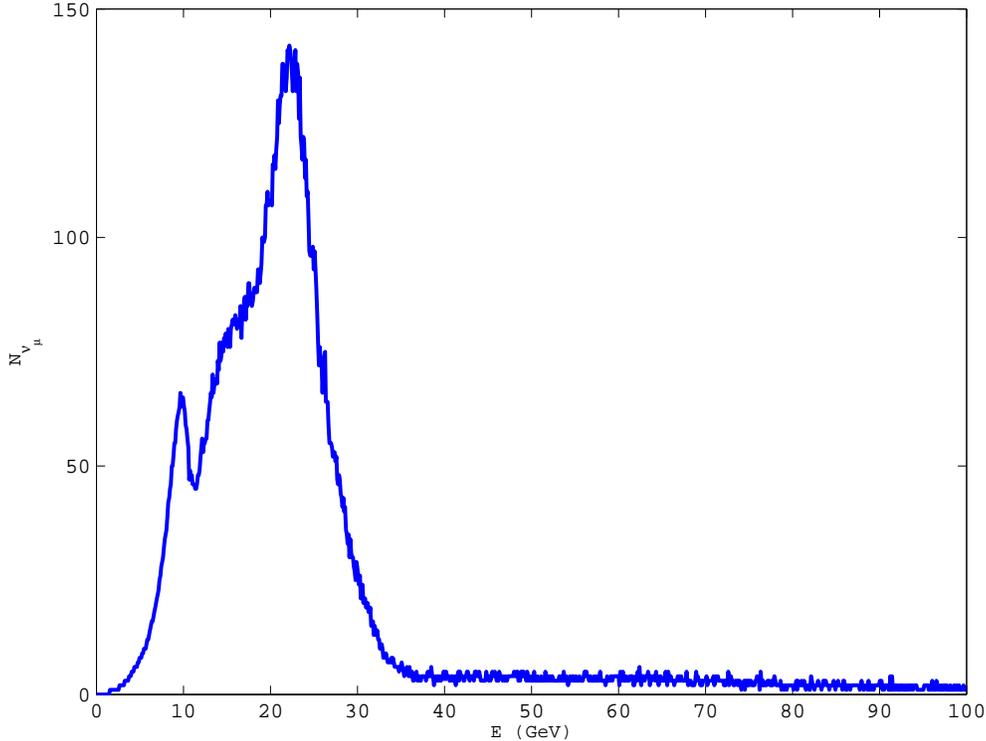,width=0.8\textwidth}
\end{center}
\caption{\it The expected CNGS neutrino beam energy
spectrum~\cite{Meddahi:2007zz}.} \label{fig:spectrum}
\end{figure}

The energy spectrum of the calculated CNGS $\nu_\mu$ flux is
reproduced in Fig.~\ref{fig:spectrum}. Its average neutrino energy
is $\sim 17$~GeV, significantly higher than that of the NuMI beam.
Since the CNGS baseline is almost identical with that the NuMI beam,
this gives some advantage to OPERA, assuming that it can attain
similar or better timing properties. We also recall that the CNGS
beam is produced by extracting the SPS beam during spills of length
$10.5~\mu$s (10500~ns). Within each spill, the beam is extracted in
2100 bunches separated by 5~ns. Each individual spill has a
$4-\sigma$ duration of 2~ns, corresponding to a Gaussian RMS width
of 0.25~ns \cite{Meddahi:2007zz}.

\subsection{Spill Analysis}

We introduce a `slicing estimator', based on the fact that if some
energy-dependent time delay is encoded into the time structure of
the spill by propagation of the neutrinos before detection, one
should observe a systematic increase in the overall time delay of
events as their energies grow. Therefore, we propose cutting the
energy spectrum of the neutrino beam into a number of energy slices,
and searching for a systematic delay in the mean arrival times of
the events belonging to different energy slices that increases with
the average energy of the slice.

\begin{figure} [htb]
\begin{center}
\epsfig{file=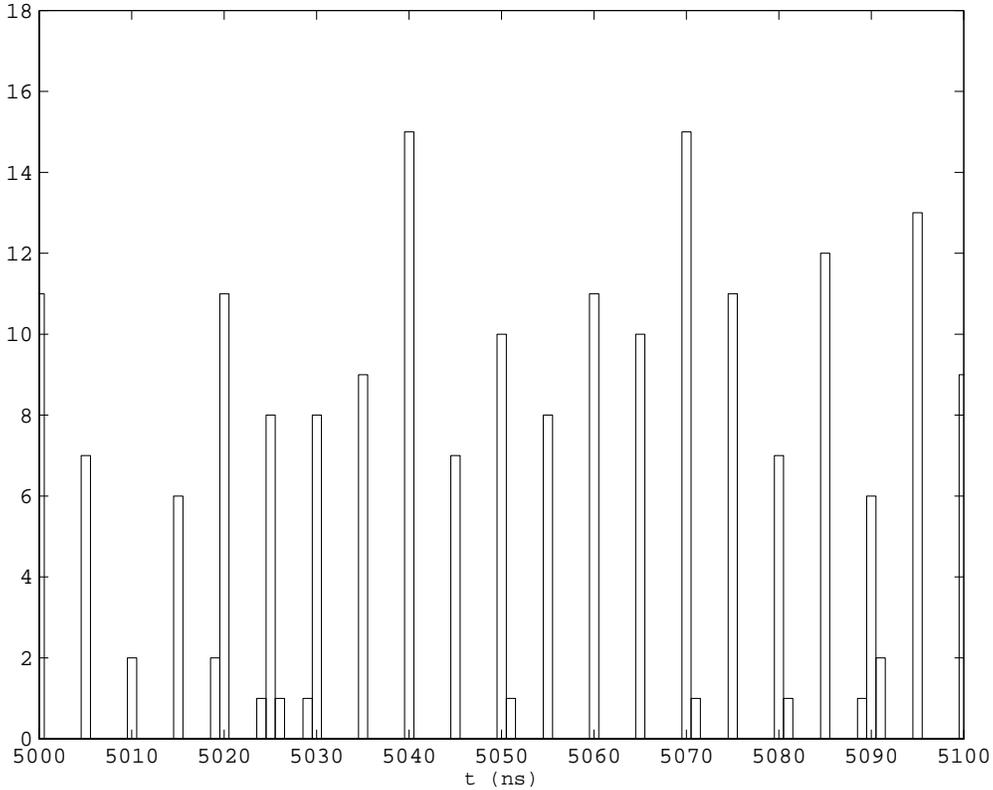,width=0.8\textwidth}
\end{center}
\caption{\it A superposition of the production times of neutrinos in
CNGS spills reflects the bunch structure of the CNGS
beam~\cite{Meddahi:2007zz}.} \label{fig:comb}
\end{figure}

In order to illustrate this idea, we perform a simple exercise
simulating the sensitivity of the slicing estimator for a time delay
depending linearly on the neutrino energy: $\Delta t=\tau E$,
assuming $\approx 2 \times 10^4$ charged-current events, as are
expected to be observed in the 1.8 kton OPERA detector over 5 years of
exposure time to the CNGS beam. We envisage superposing all the CNGS
spills with a relative timing error $\delta t$. Since each spill has
2100 bunches, we expect about 10 events on average due to each set
of superposed bunches. As a starting-point, before incorporating the
relative timing error, the timing of each event has been smeared
using a Gaussian distribution with standard deviation 0.25~ns,
reflecting the bunch spread. We display in Fig.~\ref{fig:comb} a
sample of events in our simulation, before incorporating the
relative timing error and any delay in propagation. The 5~ns
internal time structure of the spill is clearly visible.

\begin{figure} [thb]
\begin{center}
\epsfig{file=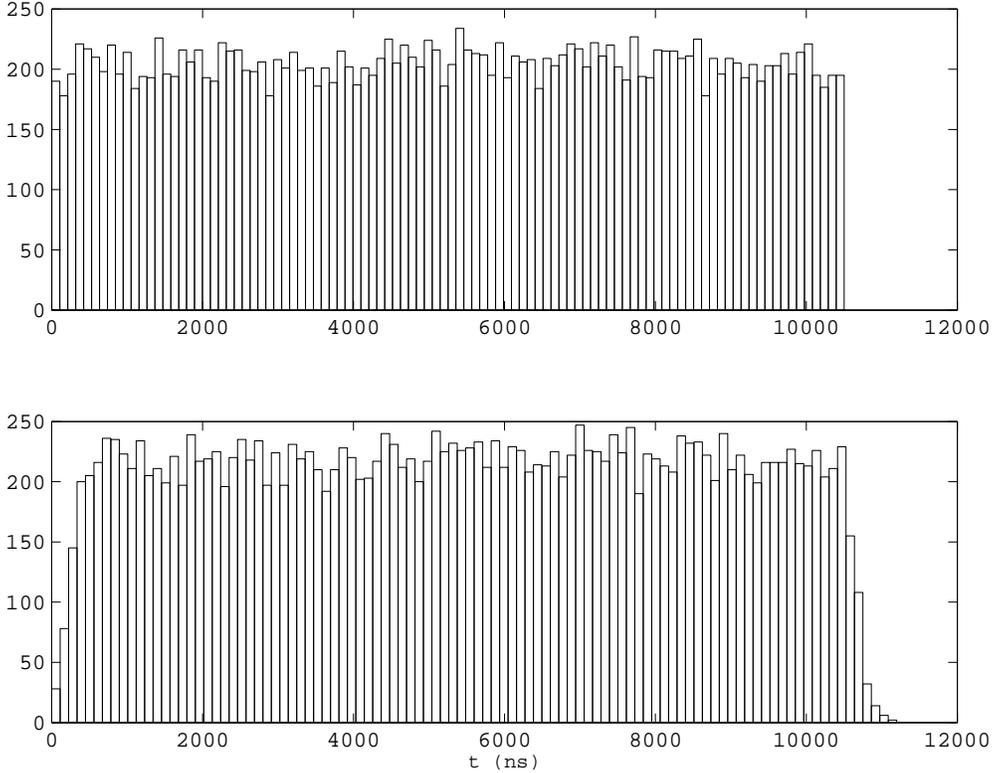,width=0.8\textwidth}
\end{center}
\caption{\it The time structure of events in the CNGS beam,
including a 100~ns timing uncertainty without (upper panel) Lorentz
violation in neutrino propagation, and (lower panel) with a linearly
energy-dependent time delay during neutrino propagation at the level
of $\tau =5$~ns/GeV.} \label{fig:smear}
\end{figure}

We now incorporate the uncertainty in the relative timing of the
bunch extraction and the detection of an event in the detector. The
overall uncertainty has three components: an uncertainty in the
extraction time relative to a standard clock at CERN, an uncertainty
in the relative timing of clocks at CERN and the LNGS provided by
the GPS system, and the uncertainty in the detector timing relative
to a standard clock in the LNGS. With the current beam
instrumentation, implementation of GPS and detector resolution, it
is expected that this will be similar to that achieved by MINOS in
the NuMI beam, namely $\sim 100$~ns. Such a timing error renders
essentially invisible the internal bunch structure of the CNGS
spill, which looks indistinguishable from a uniform distribution
generated with the same statistics, as shown in the upper panel of
Fig.~\ref{fig:smear}.

We next demonstrate in the lower panel of Fig.~\ref{fig:smear} the
effect of a time delay during neutrino propagation at the level of
$\tau_l =5$~ns/GeV, as would occur if $M_{\nu QG1} = 4.8\times 10^5\ {\rm GeV}$.
This would correspond to  a total delay $\sim 100$~ns at
the average energy of the CNGS neutrino beam. We see clearly its
smearing effect at the beginning and end of the spill, due to the
later arrivals of the more energetic neutrinos. Our `slicing
estimator' aims to quantify this effect.

\begin{figure} [thb]
\begin{center}
\epsfig{file=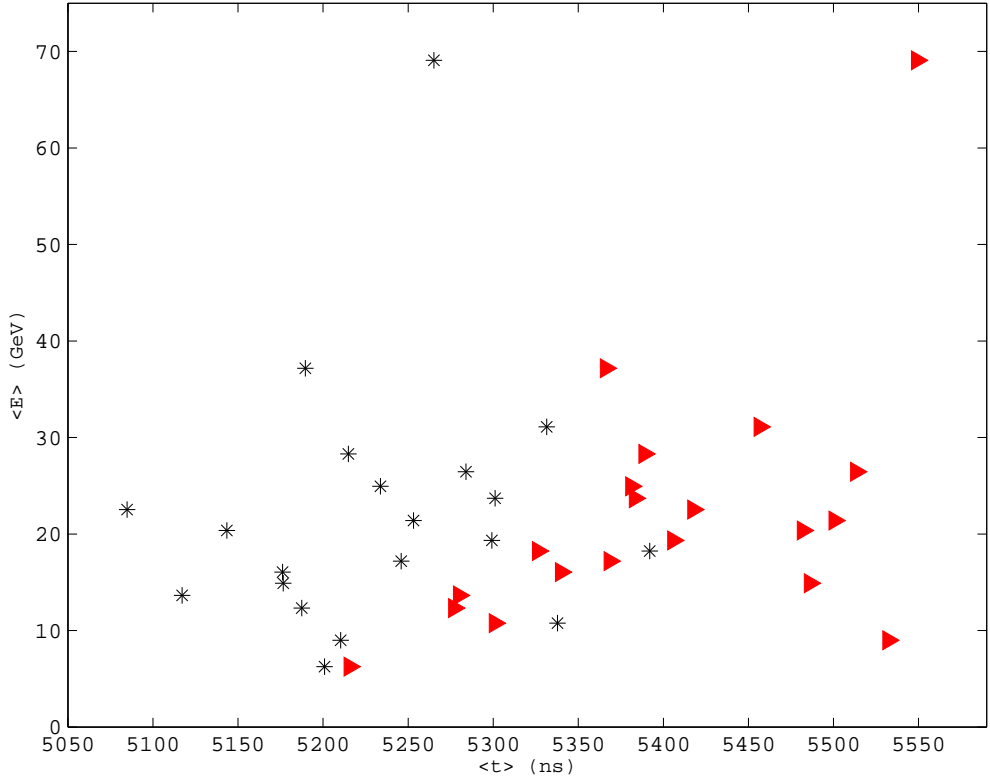,width=0.8\textwidth}
\end{center}
\caption{\it The mean arrival times of 1000-event slices with
increasing energies without Lorentz violation in the neutrino
propagation (asterisks) and with the effect of a time delay during
neutrino propagation at the level of $\tau =5$~ns/GeV (triangles). The latter
corresponds to $M_{\nu QG1} = 4.8\times 10^5\ {\rm GeV}$. One particular simulation 
of the OPERA experiment is shown: others are similar, exhibiting the
expected statistical fluctuations.}
\label{fig:slice}
\end{figure}

We smear the events with an energy resolution of 20\%, and then cut
the sample into slices of about 1000 events each with increasing
energies. The asterisks in Fig.~\ref{fig:slice} show the mean
arrival times of each slice, relative to the mean time of the
superposed spills, using one particular smearing of the timing with
a Gaussian error $\delta t = 100$~ns. The triangles in
Fig.~\ref{fig:slice}, on the other hand, show the mean arrival times
of events in each energy slice if the propagation delays caused by an
assumed value of $\tau =5$~ns/GeV are included. We see clear
differences between the asterisks and the red and triangles, that increase
with the energies of the slices.

\begin{figure} [thb]
\begin{center}
\epsfig{file=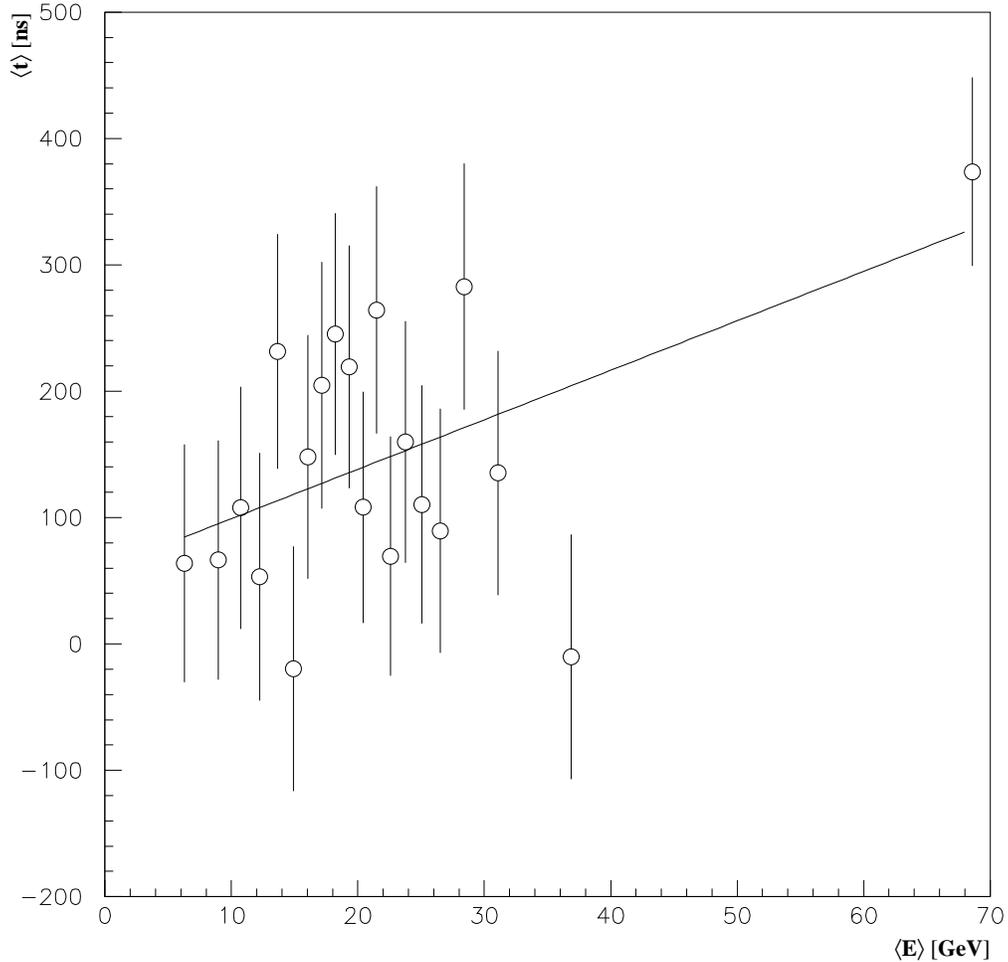,width=0.8\textwidth}
\end{center}
\caption{\it The measured shifts in the average arrival times of
neutrinos in 1000-event slices with increasing energies, assuming a
time delay during neutrino propagation at the level of $\tau
=5$~ns/GeV.} \label{fig:shift}
\end{figure}

\begin{figure} [thb]
\begin{center}
\epsfig{file=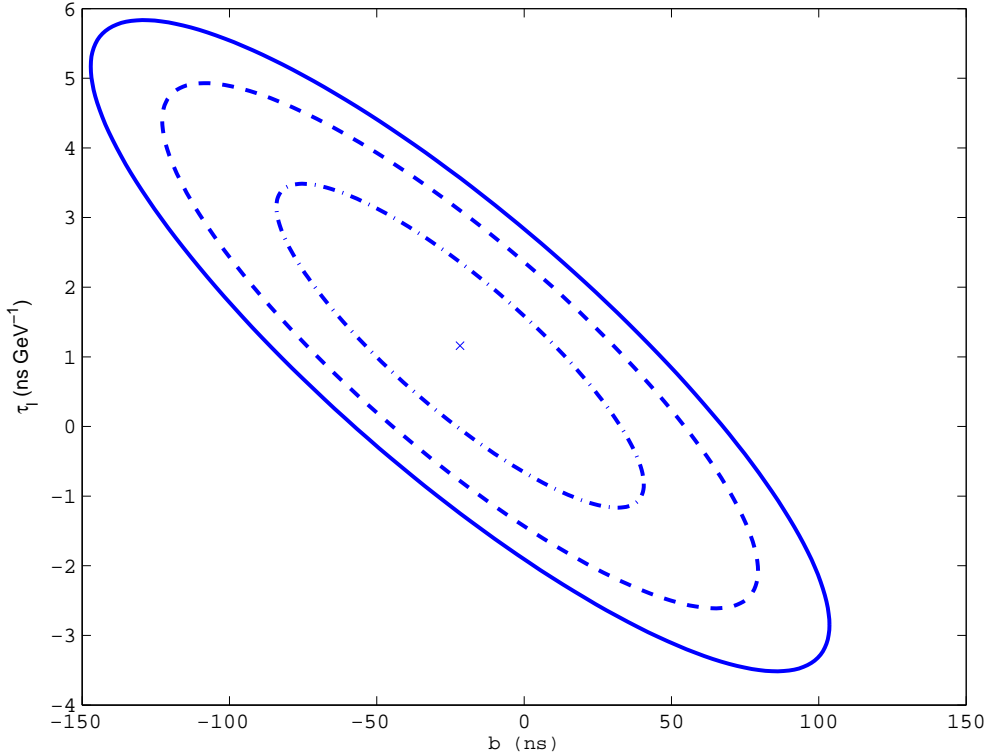,width=0.8\textwidth}
\end{center}
\caption{\it The 68\% (dashed dotted line), 95\% (dashed line) and
99\% (solid line) sensitivity contours for the case of linear
energy-dependent fit~(\ref{linfit}).} \label{fig:ellipses}
\end{figure}

By making many realizations of the event sample with the Gaussian
$\delta t = 100$~ns smearing, one can understand the significance of
the shifts in the mean positions of the slices. Fig.~\ref{fig:shift}
shows the energy dependence of the shifts in the mean timings of the
slices of 1000 events with a delay $\tau_l =5$~ns/GeV encoded. These
points may be fitted to a straight line
\begin{equation}
\label{linfit} \Delta\langle t\rangle =\tau_l\langle E\rangle +b.
\end{equation}
In general, when choosing the number of events for each slice, one
has to strike a balance between the statistics of each subsample
(which determines the precision of the determination of the mean
arrival time of each slice), and the number of subsamples to be
included in the fit. We choose the statistics of each slice so as to
give comparable error bars for each energy bin. The propagation
effect of interest to us is reflected in the slope $\tau_l$, while
the intercept is an overall shift that has no physical significance.
The sensitivity of the experiment to linear Lorentz violation at,
say, the 95\% confidence level (C.L.) may be estimated by finding
the value of the parameter $\tau_l$ which yields a fitted slope
parameter that differs from a horizontal line ($\tau_l=0$) by
$1.95\sigma$ or more. We show in Fig.~\ref{fig:ellipses} the
confidence contours corresponding to 68\% , 95\% and 99\%
sensitivity levels in the $(b, \tau_l)$ plane. From the upper
(lower) edge of the corresponding ellipse one obtains $\tau_{\rm
l95\%}=4.9(2.6)\ {\rm ns/GeV}$ at the 95\% C.L. for the subluminal
(superluminal) propagation schemes, corresponding via
\begin{equation}
\label{linscale} M_{\nu QG1}=\frac{L_{\rm CNGS}}{c}\tau_{\rm
l}^{-1}=2.4\times 10^6\left(\frac{ns~GeV^{-1}}{\tau_{\rm l}}\right)
~{\rm GeV}
\end{equation}
to values of the linear Lorentz-violating scale $M_{\nu QG1} =
4.9(9.2)\times 10^5\ {\rm GeV}$ for the subluminal (superluminal)
case, yielding a mean sensitivity~\footnote{Since the CNGS spill is
in principle time symmetric, the estimated sensitivities for sub-
and superluminal propagation should be the same. The difference
between these numbers reflects the finite size of the simulated
sample. Here and subsequently we quote the means of our sum and
superluminal limits as estimates of the CNGS sensitivity.} to
$M_{\nu QG1}\simeq 7\times 10^5\ {\rm GeV}$.
%\begin{figure} [thb]
%\begin{center}
%\epsfig{file=sens_100_100.eps,width=0.8\textwidth}
%\end{center}
%\caption{\it As for Fig.~\protect\ref{fig:shift} in the case where linear
%Lorentz violation can
%be distinguished at the 3-$\sigma$ level, i.e., with $\tau_{3\sigma}=0.144\
%{\rm ns/GeV}$.}
%\label{fig:3sigma}
%\end{figure}
It is important to note that the slope and intercept are
anticorrelated in such a fit, as shown in Fig.~\ref{fig:ellipses}.
Our conservative estimate of the limits corresponds to the upper
(lower) edges of the ellipse.

If the velocity of the neutrino depends quadratically on the energy
of the neutrino, the slices should obey a parabolic fit
\begin{equation}
\label{quadrfit} \Delta\langle t\rangle =\tau_q\langle E\rangle^2
+c.
\end{equation}
Here the propagation effect of interest is parameterized by
$\tau_q$, while an overall shift is reflected in the constant $c$.
\begin{figure} [thb]
\begin{center}
\epsfig{file=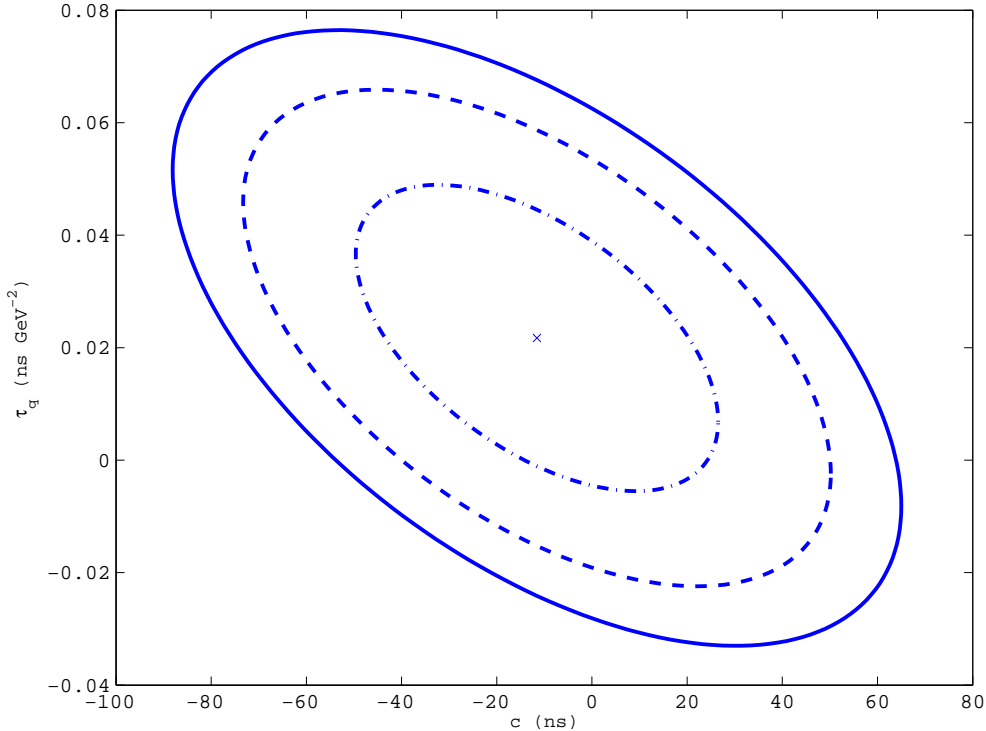,width=0.8\textwidth}
\end{center}
\caption{\it The same as in Fig.~\ref{fig:ellipses} calculated for
the sensitivity of the quadratically energy-dependent
fit~(\ref{quadrfit}).} \label{fig:ellipses2}
\end{figure}
The sensitivity contours at 68\% , 95\% and 99\% CL are presented in
Fig.~\ref{fig:ellipses2}. According to the formula
\begin{equation}
\label{quadscale} M_{\nu QG2}=\sqrt{\frac{L_{\rm CNGS}}{c}\tau_{\rm
q}^{-1}}=1.6\times 10^3\sqrt{\frac{ns~GeV^{-2}}{\tau_{\rm q}}}~{\rm
GeV},
\end{equation}
after substituting $\tau_{\rm q95\%}=0.066(0.022)$, we obtain
$M_{\nu QG1}= 6.2(11)\times 10^3\ {\rm GeV}\simeq 8\times 10^3\ {\rm
GeV}$.

The stability of the slicing estimator has been checked by
generating several data sets that have linear or quadratic
dispersion effects artificially encoded. To test our level of
sensitivity, we chose the Lorentz-violating parameters to be close
to our estimations of the levels of sensitivities in the case of
where dispersion effects are absent. The encoded values have been
recovered for the linear (\ref{linfit}) and quadratic
(\ref{quadrfit}) fits to slices containing the same numbers of
events. Slight variations in the numbers of events in the individual
slices do not change substantially the levels of sensitivity
estimated for 1000-event bins. Another check has been performed
using the minimal dispersion method described in Section 2.2.1. This
method has been applied to the whole sample of about $2\times 10^5$
events expected to occur in the rock upstream of the OPERA detector,
and results very similar to those of the slicing estimator have been
obtained. Although the whole data sample is very rich statistically,
the time distribution, given the 100~ns time uncertanty assumed in
the current analysis, is still featureless apart the edges of the
spill~\footnote{For this reason, the ECF technique described in
Section 2 is inapplicable.}.

Another check has been made analyzing the distortion of the shape of
the spill at its edges. For this purpose, we analyze two independent
histograms of the type shown in Fig.~\ref{fig:smear}. One of the
histograms is treated as a reference, while the other was shifted by
introducing a time delay $\tau_{\rm l(q)}$ for every event,
corresponding to the linear (quadratic) propagation scheme. The
shifted histogram has been compared to the reference (unshifted)
histogram, and the parameter $\tau_{\rm l(q)}$ increased until the
difference between two histograms reaches the 95\% C.L.
\begin{figure}[thb]
\begin{center}
\epsfig{file=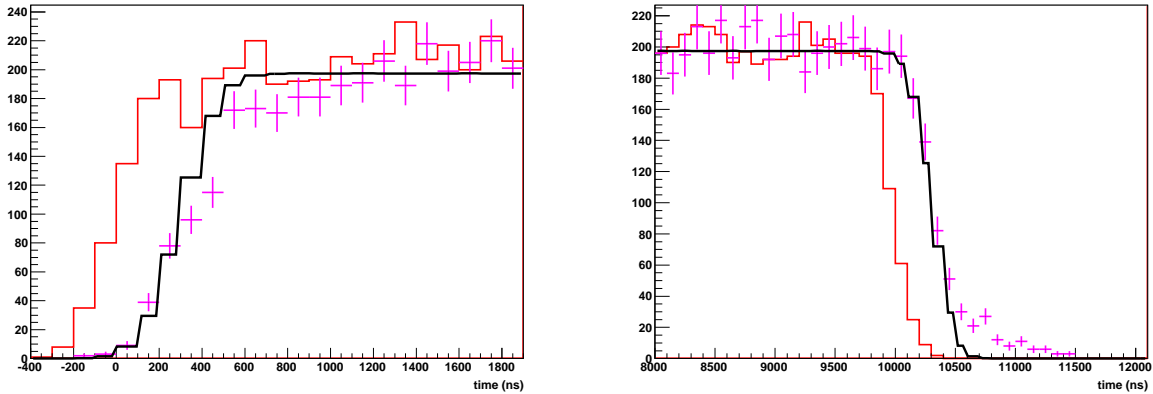,width=1.0\textwidth}
\end{center}
\caption{\it The left (right) spill edges fitted using 20000
detector events for scenario with a linear energy dependence of the
neutrino velocity. The solid red line is the reference histogram,
while the points represent the shifted data. The solid black line
represents the probability distribution function.} \label{corr_cngs}
\end{figure}
We find that this edge-fitting method has a factor 5 less
sensitivity than that obtained earlier with the slicing estimator or
the MD method.

We recall that the OPERA detector may also be used to measure the
arrival times of muons from $2 \times 10^5$ neutrino events in the
rock upstream of the detector. Information on the neutrino energy is
missing in this measurement. Therefore, one cannot employ methods
involving time-energy correlation information such as the slicing
estimator. Methods requiring an energy-dependent time shift of the
data, like the MD method, are also not applicable in this case,
again because events in the rock do not have measured energies.
Nevertheless, one can use methods that compare overall the time
shift of the simulated data to the measured time distribution of the
rock events. In this spirit, applying to the $2\times 10^5$ expected
rock events the edge-fitting procedure described in the previous
paragraph, we find a sensitivity to $M_{\nu QG1} \approx 2.4\times
10^6\ {\rm GeV}$, about three times better than previously, for the
sensitivity to linear energy dependence, and the same level of
sensitivity for the quadratic energy dependence.

One can also modify the MD method for analyzing rock events. Namely,
one could generate the reference spill and introduce an
energy-dependent shift via the parameter $\tau_{\rm l(q)}$ so as to
make the dispersion of the shifted reference spill match as closely
as possible the dispersion of the events measured in the rocks.
However, due to statistical uncertainties the dispersion of each
reference spill will be different to the dispersion of the rock events.
If this uncertainty is much less than the increase in the dispersion
of the rock events due to Lorentz violation then this method can be
used. However, this limits the sensitivity to $M_{\nu QG1} \simeq
3\times 10^5\ {\rm GeV}$ for the linear propagation scheme, which is
not as sensitive to other methods we have described above. From the
other side the sensitivity of this modified MD method approachs to
$M_{\nu QG2} \simeq 7\times 10^3\ {\rm GeV}$, which is similar to
the slicing estimator.

\subsection{Bunch Analysis}

We now explore the additional sensitivity that OPERA could obtain if
it could achieve a correlation between the SPS RF bunch structure
and the detector at the nanosecond level. Sub-ns resolution could be
obtained in OPERA with the help of additional specialized timing
detectors such as TOF hodoscopes~\footnote{We point out that it is
sufficient to refer all measured far times to a well-defined plane
perpendicular to the beam axis.}. However, synchronizing the SPS and
OPERA clocks with such a precision over a period of 5~years is a
challenging task. With the new IEEE Standard Precision Time Protocol
(PTP) IEEE1588~\cite{IEEE1588} it is possible achieve time
synchronization in the range of 100~ns on an Ethernet network but
not better; GPS clock synchronization at the ns level is also highly
demanding.  Standard `One-Way' GPS techniques~\cite{OWGPS} can reach
a precision of $\sim 20$~ns at best. Devices known as GPS
disciplined oscillations (GPSDO)~\cite{GPSDO}, containing
high-quality temperature-controlled local oscillators, steered to
agree with the onboard oscillators of the GPS satellites,  can
provide ultra-precise standard frequencies that could reproduce the
CERN RF frequency. A more elegant but less standard method is called
`Common-View' GPS~\cite{OWGPS}: in this case two clocks (e.g., one
at CERN and the other at LNGS) view simultaneously the same GPS
satellite, thereby cancelling out the common errors (e.g., the
satellite's local clock). It has been shown that the data recorded
by the two GPS receivers can be processed offline to provide a
timing uncertainty $\lesssim 5$~ns. Finally it has been shown that
`Carrier-Phase' GPS measurements~\cite{CPGPS}, which use the carrier
frequencies instead of the codes transmitted by the satellites, can
achieve synchronization of clocks with uncertainties $\sim 0.5$~ns
at the cost of extensive post processing. Turning to ground based
solutions, the most precise atomic clock (the NIST-F1 used to define
the UTC) has a long-term accuracy of $5\times 10^{-16}$ or $\sim
75$~ns over 5 years. It would therefore not be sufficient to bring
two {\it a priori} synchronized clocks to the near and far locations
to define the arrival times with the required long-term stability.
Alternatively, next-generation accelerators, e.g., free electron
lasers such as XFELs that aim to generate X-ray pulses with pulse
durations down to tens of femtoseconds, will meet the challenge of
finding new methods of ultra-stable timing stabilization,
synchronization and distribution over several kilometres. These
systems will most likely rely on optical timing synchronization. We
can therefore imagine a phase-locked loop RF oscillator located at
the far location remotely locked to the SPS RF system. These two
systems would be connected and locked via stabilized optical fibre
links~\footnote{We note that the temperature dependence of the
refractive index of an optical fibre is typically $10^{-6}$/K, which
corresponds to a drift of 5~ns for 1000~km and a temperature
stability of 1$^\circ$C.}. To conclude, a combination of space- or
ground-based solutions could probably provide the possible
synchronization of the CNGS and OPERA clocks, and allow for
systematic cross-checks to be performed.

\begin{figure} [thb]
\begin{center}
\epsfig{file=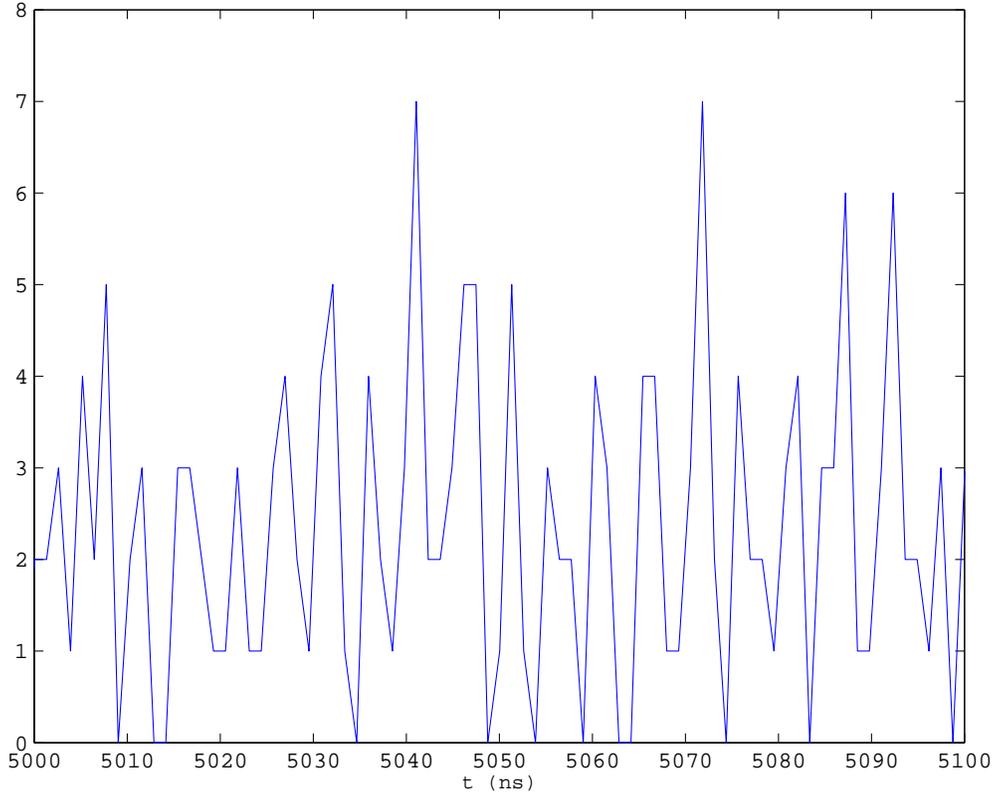,width=0.8\textwidth}
\end{center}
\caption{\it A particular realization of the bunch structure with
$\approx 1$~ns relative time uncertainty incorporated. The histogram
is binned with a resolution sutable for resolving the bunch
structure.} \label{fig:batch_smeared}
\end{figure}

We now discuss how the sensitivity of the previous analysis could be
improved by taking into account the 5-ns bunch structure of the CNGS
spills. In Fig.~\ref{fig:batch_smeared} we present one particular
realization of a sample of simulated events which incorporates a
relative timing error of 1ns.
Although the periodic bunch structure survives, the signal itself
represents a time series with a relatively low signal-to-noise
ratio. The latter implies that the proper deconvolution to extract
isolated features cannot be made. In the other words, there is a
problem in fitting the fine structure of the signal with an
analytical function. Such a situation has been widely investigated
and applied to the temporal profiles of gamma gamma ray bursters
(GRBs)~\cite{ccf_grb}. We therefore apply a cross correlation
function (CCF) method similar to that described in~\cite{ccf_grb}
but differing only in details of its adaptation. Namely, we
introduce the temporal correlation of two time series $A(t)$ and
$B(t+\tau_{l(q)})$
\begin{equation}
\label{ccf} {\rm CCF(\tau_{l(q)})}=\frac{\langle (A(t)-\bar
A(t))(B(t-\tau_{l}E^{l})-\bar B(t-\tau_{l}E^{l}))
\rangle}{\sigma_{A(t)}\sigma_{B(t-\tau_{l}E^{l})}},
\end{equation}
where $A(t)$ is a Monte Carlo simulation of the events with no
dispersion effects, $B(t-\tau_{l}E^{l})$ is the simulated data which
has the time shift required to invert the effect of the
energy-dependent dispersion, $\bar A(t)$ and $\bar
B(t-\tau_{l}E^{l})$ are the mean values of the corresponding time
series, and $\sigma_{A(t)}$ and $\sigma_{B(t-\tau_{l}E^{l})}$ are
the standard deviations from these mean values. We average over
several Monte Carlo simulations to include the statistical
uncertainties as well as performing time and energy smearing due to
the uncertainty in these measurements.

\begin{figure} [thb]
\begin{center}
\epsfig{file=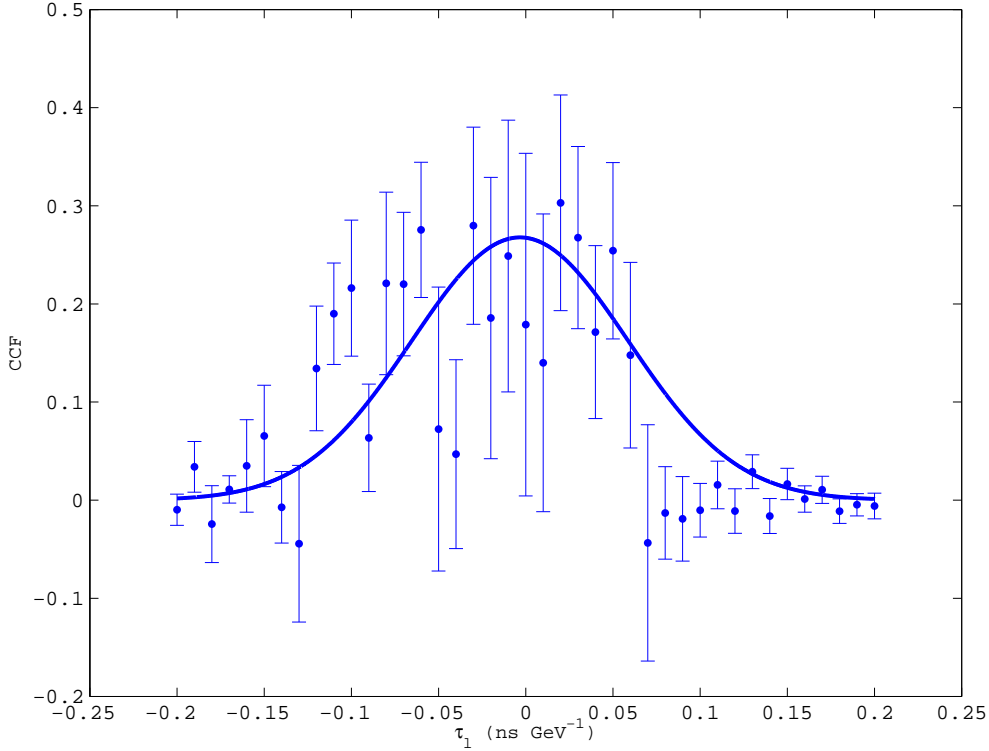,width=0.8\textwidth}
\end{center}
\caption{\it The Gaussian fit to the CCF calculated for the case of
a linear energy dependence with time smearing $\approx 1$~ns. }
\label{fig:ccf_lin}
\end{figure}

We calculate the ${\rm CCF(\tau_{l(q)})}$ as a function of
$\tau_{l}$ and find its maximum value. The value of $\tau_{l}$ which
maximizes the CCF is an estimate of the true value of $\tau_{l}$. To
find this estimate we fit a Gaussian to the peak of the resulting
CCF function shown in Fig.~\ref{fig:ccf_lin}.
Each realization produced an independent measurement of the CCF at a
given value of the shift parameter. The process of iteration for
every value of the shift parameter in Fig.~\ref{fig:ccf_lin} was
repeated until the resulting distribution approached a normal
distribution, which typically took about 100 runs. Using these
normal distributions the values and the standard deviations (error
bars) presented in Fig.~\ref{fig:ccf_lin} have been calculated.

\begin{figure} [thb]
\begin{center}
\epsfig{file=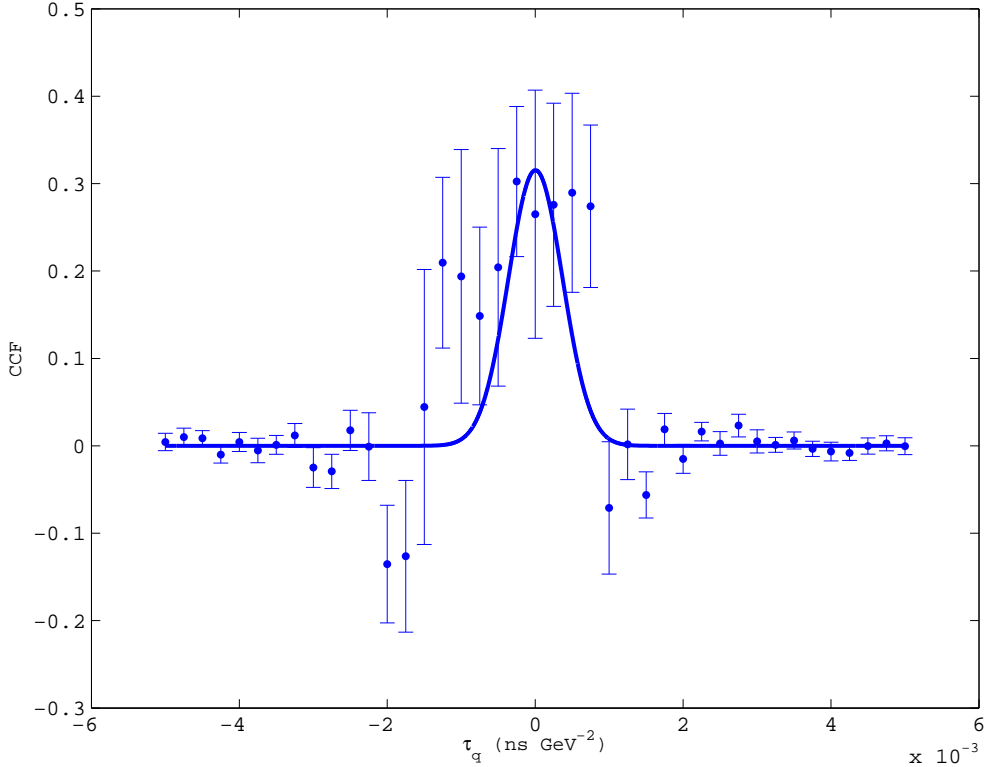,width=0.8\textwidth}
\end{center}
\caption{\it The same as in Fig.~\ref{fig:ccf_lin} for the quadratic
energy dependence.} \label{fig:ccf_quadr}
\end{figure}

The sensitivity of the CCF can then be estimated by the precision of
the position of the maximum for the Gaussian fit in
Fig.~\ref{fig:ccf_lin}. For the case of linear energy dispersion,
the maximum of the CCF is found at $\tau_{\rm l}^{\rm max}=-0.033\pm
0.036\ {\rm ns/GeV}$ if no time shift encoded in the simulated data.
For superluminal propagation, when $\tau_{\rm l}>0$, one can
estimate $\tau_{\rm l95\%}^{\rm su}=0.037\ {\rm ns/GeV}$, which
corresponds via (\ref{linscale})  to $M_{\nu QG1} \approx 6.6\times
10^7\ {\rm GeV}$. For the subluminal case, one obtains $\tau_{\rm
l95\%}^{\rm sb}=0.1{\rm ns/GeV}$, which corresponds to $M_{\nu QG1}
\approx 2.4\times 10^7\ {\rm GeV}$. The same CCF procedure may also
be applied to the quadratic case, as shown in
Fig.~\ref{fig:ccf_quadr}.
The limits deduced from the fit Fig.~\ref{fig:ccf_quadr} are $M_{\nu
QG2}= 3.6(4.9) \times 10^{4}\ {\rm GeV}\simeq 4\times 10^{4}\ {\rm
GeV}$.

\begin{figure} [thb]
\begin{center}
\epsfig{file=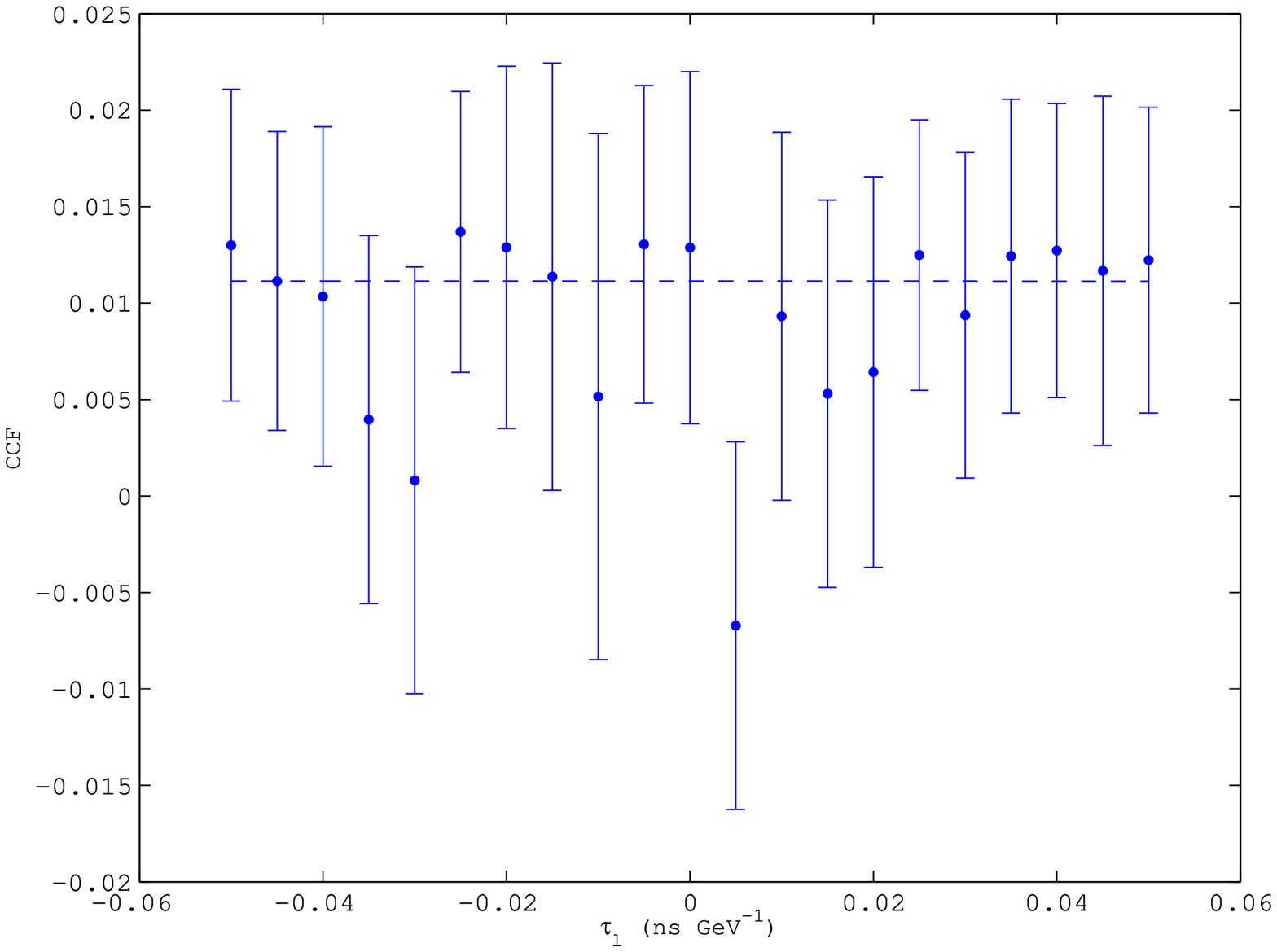,width=0.8\textwidth}
\end{center}
\caption{\it The profile of the CCF calculated with a 2~ns time
resolution for the case of linear energy dependence in neutrino
propagation.} \label{fig:ccf_flat}
\end{figure}

Repeating the CCF procedure for a time resolution above 2~ns, one
observes no maximum correlation in a reasonable range of the shift
parameter, as seen in Fig.~\ref{fig:ccf_flat}.
From this one can conclude that the bunch structure degenerates into
an essentially uniform distribution as soon as the time resolution
becomes bigger than $\approx 2$~ns, in which case the slicing
estimator described in the previous subsection should be applied.

If the same time resolution $\sim 1$~ns can be attained
for events occurring in the rock upstream from the OPERA detector,
the CCF method can also be used to analyze these data, which should
amount to some $2 \times 10^5$ events.
\begin{figure} [thb]
\begin{center}
\epsfig{file=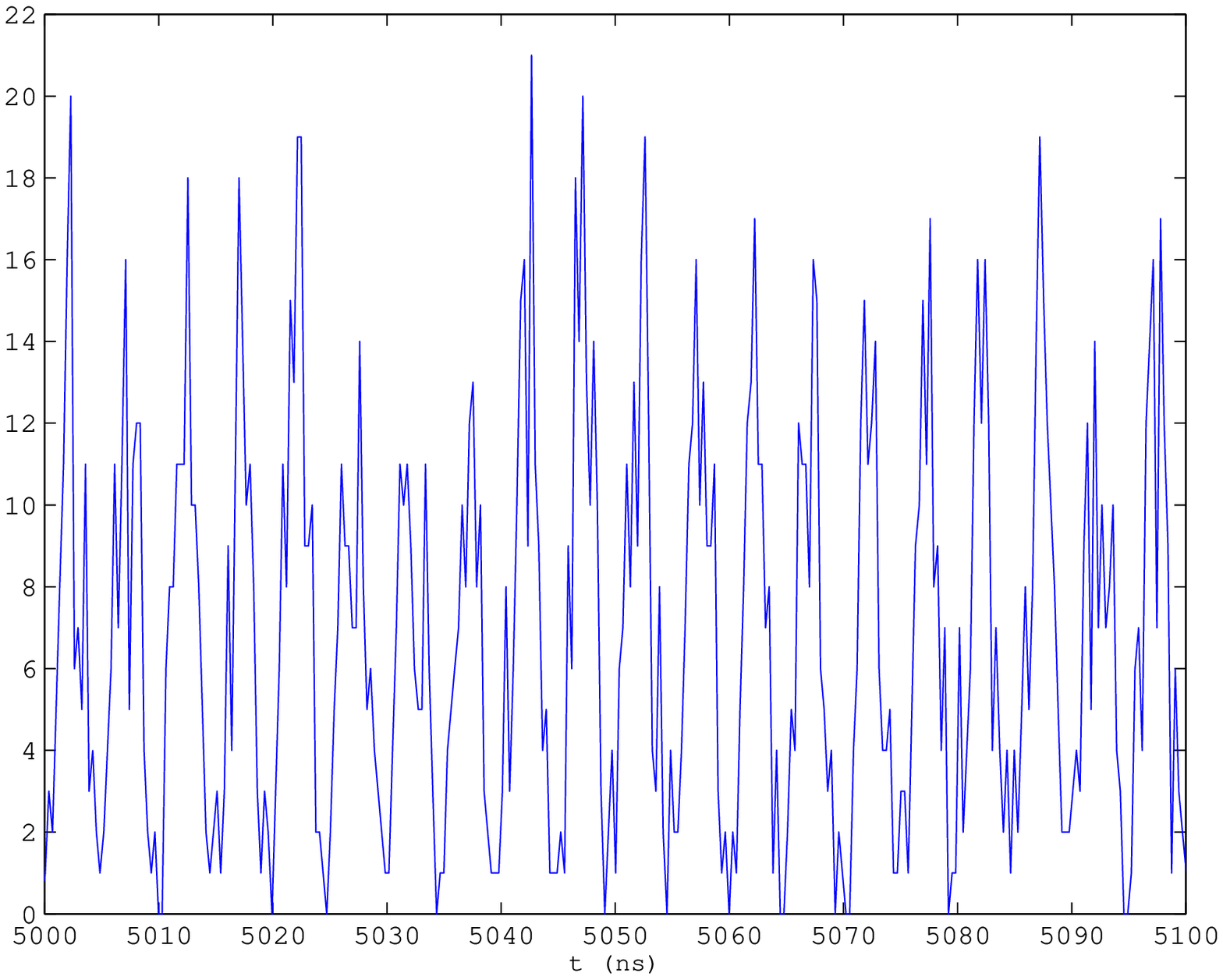,width=0.8\textwidth}
\end{center}
\caption{\it A simulated realization of the bunch structure for rock
events, incorporating a timing uncertainty $\approx 1$~ns. The
histogram is binned with a resolution suitable for resolving the
bunch structure.} \label{fig:batch_smeared1}
\end{figure}
We see in Fig.~\ref{fig:batch_smeared1} that the bunch structure of
the rock events is clearly visible if a time resolution $\approx
1$~ns is achieved, despite the fact that the energies of the
neutrinos colliding in the rock cannot be determined.
 \begin{figure} [thb]
\begin{center}
\epsfig{file=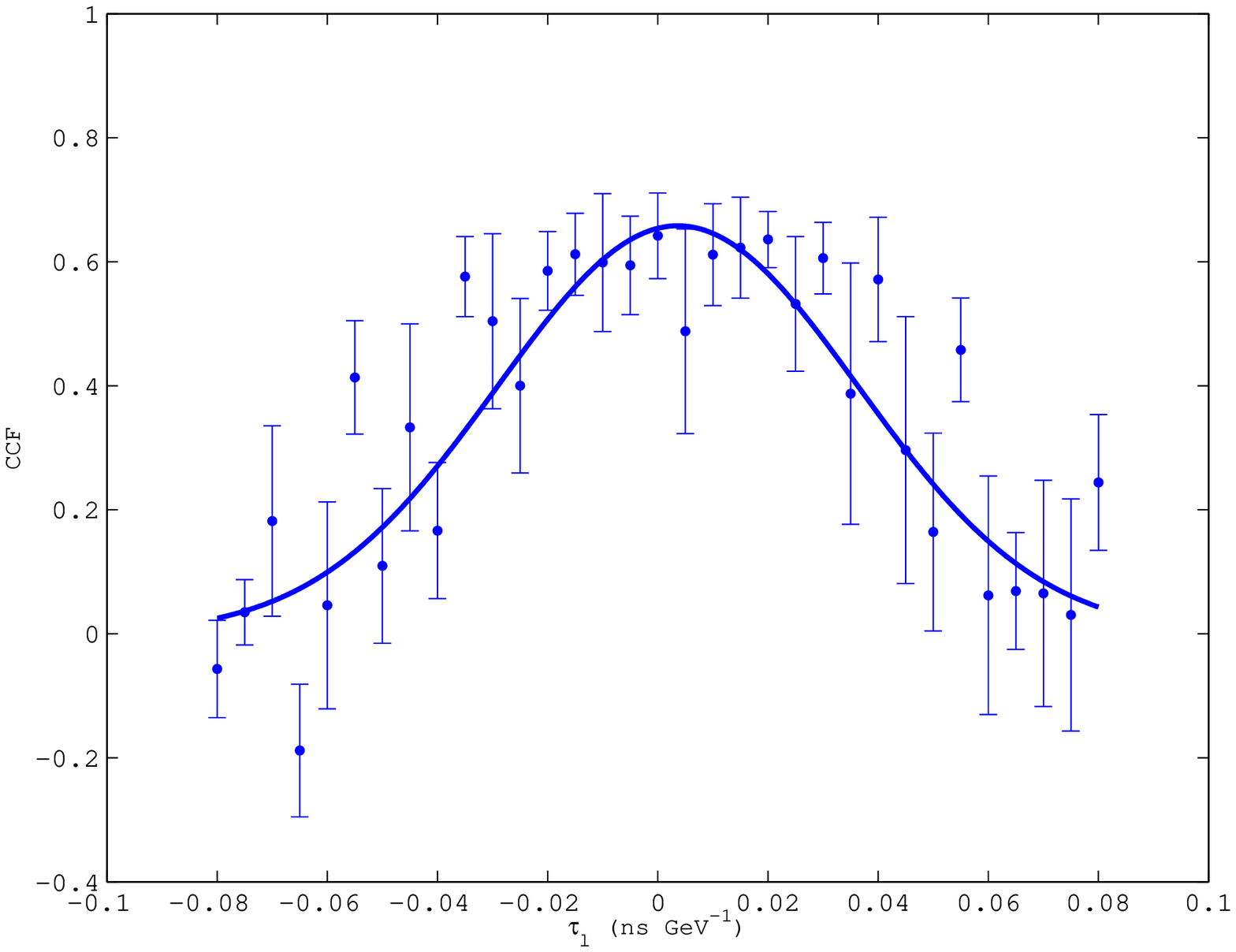,width=0.8\textwidth}
\end{center}
\caption{\it The CCF for rock events with time resolution $\approx
1$~ns in the case of linear energy dependence, compared with a
Gaussian fit.} \label{fig:ccf_rock}
\end{figure}
The CCF calculated for the rock events is presented in
Fig.~\ref{fig:ccf_rock}, together with a Gaussian fit. The
sensitivities to Lorentz violation now attain the levels of $M_{\nu
QG1}= 4.3(3.2)\times 10^8\ {\rm GeV}\simeq 4\times 10^8\ {\rm GeV}$
for the linear case, and $M_{\nu QG2}= 8.8(4.3) \times 10^{5}\ {\rm
GeV}\simeq 7 \times 10^{5}\ {\rm GeV}$ for the quadratic case. The
sensitivity in the quadratic case is significantly better than the
sensitivity estimated for a possible future galactic supernova.

%%%%%%%%%%%%%%%%%%%%%%%%%%%%%%%%%%%%%%%%%%%%%%%%%%%%%%%%%%%%%%%%%%%%%%%%%%%%%%%%%%%%%%%
\section{Conclusions}
%%%%%%%%%%%%%%%%%%%%%%%%%%%%%%%%%%%%%%%%%%%%%%%%%%%%%%%%%%%%%%%%%%%%%%%%%%%%%%%%%%%%%%%

We find from the SN1987a data lower limits on the scale of linear
Lorentz violation in the neutrino sector, namely $M_{\nu QG1}
> 2.68 \times 10^{10}$~GeV and $M_{\nu QG1}
> 2.51 \times 10^{10}$~GeV at the $95\%$ C.L. in the subluminal and
superluminal cases respectively. The corresponding limits for the
quadratic models are $M_{\nu QG2} > 4.62 \times 10^{4}$~GeV and
$M_{\nu QG2} > 4.13 \times 10^{4}$~GeV at the $95\%$ C.L. in the
subluminal and superluminal cases, respectively. We have also used a
Monte Carlo simulation of a galactic supernova at 10~kpc to estimate
how accurately Lorentz violation could be probed in the future. In
such a case one would observe more events because of the larger
fiducial volume of the SK detector compared to the previous
generation of detectors, and because the next observable supernova
is likely to be inside the galaxy and hence closer than SN1987a. On
the other hand, if the next supernova is closer than SN1987a then
the energy-dependent time shift due to Lorentz violation will be
reduced, reducing also the expected sensitivity. We performed
simulations for both the normal and inverted mass hierarchies and
for both an adiabatic and a non-adiabatic atmospheric resonance. In
all scenarios it would be possible to probe Lorentz violation using
the methods decribed in this paper. We used the minimal dispersion
(MD) method and the maximal ECF method with a several energy
weightings and have shown that using the latter with a linear energy
weighting has the greatest sensitivity. Using this method we have
shown that we could place limits up to $M_{\nu QG1}> 2.2 \times
10^{11}$~GeV and $M_{\nu QG1}> 4.2 \times 10^{11}$~GeV at the $95\%$
C.L. for the subluminal and superluminal cases, respectively, for
linear models of Lorentz violation, and $M_{\nu QG2}> 2.3 \times
10^{5}$~GeV and $M_{\nu QG2}> 3.9 \times 10^{5}$~GeV at the $95\%$
C.L. for the subluminal and superluminal cases, respectively, for
quadratic models of Lorentz violation.

We have then explored the sensitivity to Lorentz
violation in neutrino propagation that could be obtained using data
from the OPERA detector in the CNGS beam. By comparison with the
result already obtained by MINOS in the NuMI beam, OPERA would
benefit from the higher energy of the CNGS beam, the larger
statistics we assume, and the possibility of exploiting the bunch
structure of the CNGS beam that we have explored. We find that,
using standard clock synchronization techniques,  the sensitivity of
the OPERA experiment would reach $M_{\nu QG1} \sim 7 \times
10^{5}$~GeV ($M_{\nu QG2} \sim 8 \times 10^{3}$~GeV) after 5 years
of nominal running. If the time structure of the SPS RF bunches
within the extracted CNGS spills of 10.5~$\mu$s could be exploited,
which would require reducing the timing uncertainty to $\sim 1$~ns,
these figures would be significantly improved to $M_{\nu QG1}\sim 5
\times 10^{7}$~GeV ($M_{\nu QG2} \sim 4 \times 10^{4}$~GeV). Using
events in the rock upstream of OPERA, and again assuming a time
resolution $\sim 1$~ns, the sensitivities to Lorentz violation could
be further improved to $M_{\nu QG1} \simeq 4\times 10^8\ {\rm GeV}$
for the linear case and $M_{\nu QG2}= \simeq 7 \times 10^{5}\ {\rm
GeV}$ for the quadratic case. While still inferior to the
sensitivity of the supernova limits in the linear case, the OPERA
rock sensitivity in the quadratic case would exceed even that
possible using data from a future galactic supernova. This and the
fact that any accelerator limit benefits from better-understood
experimental conditions would motivate the effort that would be
required to achieve nanosecond time resolution for the OPERA/CNGS
combination.

\section*{Acknowledgements}

We thank N.~E.~Mavromatos, D.~V.~Nanopoulos and E.~K.~G.~Sarkisyan for
discussions on related subjects. N.~H. and A.~S.~S. thank the CERN Theory
Division for its kind hospitality. N.~H. also thanks STFC for the Studentship
Award PPA/S/S/2004/03926, and the UniverseNet network for supporting
this research project by a Marie Curie Early Stage Research Training
Fellowship of the European Community's Sixth Framework Programme
under contract (MRTN-CT-2006-0355863-UniverseNet).


\begin{thebibliography}{100}

%astrophysical and long baseline souces
\bibitem{Strumia:2006db}
  For reviews, see:A.~Strumia and F.~Vissani,
  %``Neutrino masses and mixings and.,''
  arXiv:hep-ph/0606054.
  %%CITATION = HEP-PH/0606054;%%

\bibitem{Totsuka:1991dm}
  Y.~Totsuka,
  %``Neutrino astronomy,''
  Rept.\ Prog.\ Phys.\  {\bf 55} (1992) 377.
  %%CITATION = RPPHA,55,377;%%


%quantum decoherence

\bibitem{Barenboim:2006xt}
  G.~Barenboim, N.~E.~Mavromatos, S.~Sarkar and A.~Waldron-Lauda,
  %``Quantum decoherence and neutrino data,''
  Nucl.\ Phys.\  B {\bf 758} (2006) 90
  [arXiv:hep-ph/0603028].
  %%CITATION = NUPHA,B758,90;%%

\bibitem{CNGS-T2K}
  N.~E.~Mavromatos, A.~Meregaglia, A.~Rubbia, A.~Sakharov and S.~Sarkar,
  %``Quantum-Gravity Decoherence Effects in Neutrino Oscillations: Expected
  %Constraints From CNGS and J-PARC,''
  Phys.\ Rev.\  D {\bf 77} (2008) 053014
  [arXiv:0801.0872 [hep-ph]].
  %%CITATION = PHRVA,D77,053014;%%

\bibitem{Morgan:2004vv}
  D.~Morgan, E.~Winstanley, J.~Brunner and L.~F.~Thompson,
  %``Probing quantum decoherence in atmospheric neutrino oscillations with a
  %neutrino telescope,''
  Astropart.\ Phys.\  {\bf 25} (2006) 311
  [arXiv:astro-ph/0412618].
  %%CITATION = APHYE,25,311;%%

\bibitem{Hooper:2004xr}
  D.~Hooper, D.~Morgan and E.~Winstanley,
  %``Probing quantum decoherence with high-energy neutrinos,''
  Phys.\ Lett.\  B {\bf 609} (2005) 206
  [arXiv:hep-ph/0410094].
  %%CITATION = PHLTA,B609,206;%%

%Quantum Gravity

\bibitem{foam} G.~Amelino-Camelia, J.~R.~Ellis, N.~E.~Mavromatos and
D.~V.~Nanopoulos, Int.\ J.\ Mod.\ Phys.\ A {\bf 12} (1997) 607
 [arXiv:hep-th/9605211];
  %%CITATION = HEP-TH 9605211;%%
  J.~R.~Ellis, N.~E.~Mavromatos and D.~V.~Nanopoulos,
 Phys.\ Lett.\ B {\bf 293} (1992) 37
  [arXiv:hep-th/9207103];
  %%CITATION = HEP-TH 9207103;%%
  J.~R.~Ellis, N.~E.~Mavromatos and D.~V.~Nanopoulos,
 {\it Erice Subnucl. Phys. Series}, Vol. {\bf 31}
  1 (World Sci. 1994) [arXiv:hep-th/9311148];
  %%CITATION = HEP-TH 9311148;%%
  J. Chaos, Solitons  and Fractals, Vol. {\bf 10} (1999) 345
(eds. C. Castro and M.S. El Naschie, Elsevier Science, Pergamon
1999) [arXiv:hep-th/9805120];
  %%CITATION = HEP-TH 9805120;%%

\bibitem{gambini}
 R.~Gambini and J.~Pullin,
  Phys.\ Rev.\ D {\bf 59} (1999) 124021
  [arXiv:gr-qc/9809038];
 J.~Alfaro, H.~A.~Morales-Tecotl and L.~F.~Urrutia,
  Phys.\ Rev.\ D {\bf 65} (2002) 103509
  [arXiv:hep-th/0108061];
  %%CITATION = HEP-TH 0108061;%%
 V.~A.~Kostelecky and S.~Samuel,
  Phys.\ Rev.\ D {\bf 39} (1989) 683;
  %%CITATION = PHRVA,D39,683;%%
   G.~Amelino-Camelia,
  Int.\ J.\ Mod.\ Phys.\ D {\bf 11} (2002) 35
  [arXiv:gr-qc/0012051];
  %%CITATION = GR-QC 0012051;%%
 R.~C.~Myers and M.~Pospelov,
  Phys.\ Rev.\ Lett.\  {\bf 90} (2003) 211601
  [arXiv:hep-ph/0301124].
  %%CITATION = HEP-PH 0301124;%%

  \bibitem{amellis}
G.~Amelino-Camelia, J.~Ellis, N.~Mavromatos, D.~Nanopoulos and
S.~Sarkar, Nature {\bf 393} (1998) 763; J.~Ellis, K.~Farakos,
N.E.~Mavromatos, V.A.~Mitsou and D.V.~Nanopoulos, Ap. J. {\bf 535}
(2000) 139; J.~R.~Ellis, N.~E.~Mavromatos, D.~V.~Nanopoulos,
A.~S.~Sakharov and E.~K.~G.~Sarkisyan,
  %``Robust Limits on Lorentz Violation from Gamma-Ray Bursts,''
  Astropart.\ Phys.\  {\bf 25} (2006) 402
  [Astropart.\ Phys.\  {\bf 29} (2008) 158]
  [arXiv:astro-ph/0510172]; ibid J.~Ellis, N.~E.~Mavromatos,
  D.~V.~Nanopoulos, A.~S.~Sakharov and E.~K.~G.~Sarkisyan,
  %``Erratum (astro-ph/0510172): Robust Limits on Lorentz Violation from
  %Gamma-Ray Bursts,''
  arXiv:0712.2781 [astro-ph].
  %%CITATION = ARXIV:0712.2781;%%
  %%CITATION = APHYE,29,158;%%

\bibitem{pulsar} P.~Kaaret,
  Astron. Astrophys. {\bf 345} (1999) L32 [astro-ph/9903464].
  %%CITATION = ASTRO-PH 9903464;%%

\bibitem{Albert:2007qk}
  J.~Albert {\it et al.}  [MAGIC Collaboration],
  %``Probing Quantum Gravity using Photons from a Mkn 501 Flare Observed by
  %MAGIC,''
  arXiv:0708.2889 [astro-ph].
  %%CITATION = ARXIV:0708.2889;%%


\bibitem{equiv} J.~R.~Ellis, N.~E.~Mavromatos, D.~V.~Nanopoulos and
A.~S.~Sakharov,
  %``Space-time foam may violate the principle of equivalence,''
  Int.\ J.\ Mod.\ Phys.\  A {\bf 19} (2004) 4413
  [arXiv:gr-qc/0312044].
  %%CITATION = IMPAE,A19,4413;%%

\bibitem{refract_last} J.~Ellis, N.~E.~Mavromatos and D.~V.~Nanopoulos,
  %``Derivation of a Vacuum Refractive Index in a Stringy Space-Time Foam
  %Model,''
  arXiv:0804.3566 [hep-th].
  %%CITATION = ARXIV:0804.3566;%%
  
\bibitem{LeptonNumber} J.~R.~Ellis, N.~E.~Mavromatos, D.~V.~Nanopoulos
and G.~Volkov,
  %``Gravitational-recoil effects on fermion propagation in space-time foam,''
  Gen.\ Rel.\ Grav.\  {\bf 32}, 1777 (2000)
  [arXiv:gr-qc/9911055].
  %%CITATION = GRGVA,32,1777;%%

\bibitem{Rebel:2008th} P.~Adamson {\it et al.}  [MINOS Collaboration],
  %``Measurement of neutrino velocity with the MINOS detectors and NuMI neutrino
  %beam,''
  Phys.\ Rev.\  D {\bf 76} (2007) 072005
  [arXiv:0706.0437 [hep-ex]].
  %%CITATION = PHRVA,D76,072005;%%


%SN stuff

\bibitem{k2sn1987a}
R.~M.~Bionta {\it et al.},
%``Observation Of A Neutrino Burst In Coincidence With Supernova Sn1987a In The Large Magellanic Cloud,''
Phys.\ Rev.\ Lett.\  {\bf 58} (1987) 1494.
%%CITATION = PRLTA,58,1494;%%

\bibitem{imbsn1987a}
K.~Hirata {\it et al.}  [KAMIOKANDE-II Collaboration],
%``Observation Of A Neutrino Burst From The Supernova Sn1987a,''
Phys.\ Rev.\ Lett.\  {\bf 58} (1987) 1490.
%%CITATION = PRLTA,58,1490;%%


\bibitem{baksan1987a}
  E.~N.~Alekseev, L.~N.~Alekseeva, V.~I.~Volchenko and I.~V.~Krivosheina,
  %``POSSIBLE DETECTION OF A NEUTRINO SIGNAL ON 23 FEBRUARY 1987 AT THE BAKSAN
  %UNDERGROUND SCINTILLATION TELESCOPE OF THE INSTITUTE OF NUCLEAR RESEARCH,''
  JETP Lett.\  {\bf 45} (1987) 589
  [Pisma Zh.\ Eksp.\ Teor.\ Fiz.\  {\bf 45} (1987) 461].
  %%CITATION = ZFPRA,45,461;%%

  E.~N.~Alekseev, L.~N.~Alekseeva, I.~V.~Krivosheina and V.~I.~Volchenko,
  %``DETECTION OF THE NEUTRINO SIGNAL FROM SN1987A IN THE LMC USING THE INR
  %BAKSAN UNDERGROUND SCINTILLATION TELESCOPE,''
  Phys.\ Lett.\  B {\bf 205} (1988) 209.
  %%CITATION = PHLTA,B205,209;%%

% Volkov
\bibitem{volkov}
V.~Ammosov and G.~Volkov, 
%``Can neutrinos probe extra dimensions?,''
  arXiv:hep-ph/0008032;
  %%CITATION = HEP-PH/0008032;%%
  G.~G.~Volkov,
  %``Geometry of Majorana neutrino and new symmetries,''
  Annales Fond.\ Broglie {\bf 31} (2006) 227
  [arXiv:hep-ph/0607334].
  %%CITATION = AFLBD,31,227;%%
  
%CNGS

\bibitem{Meddahi:2007zz}
  M.~Meddahi {\it et al.},
  %``Cern Neutrinos To Gran Sasso -- Cngs: Results From Commissioning,''
{\it In the Proceedings of Particle Accelerator Conference (PAC 07),
Albuquerque, New Mexico, 25-29 Jun 2007, pp 692}.
  %%CITATION = CONFP,C070625,692;%%


%water cherenkov
\bibitem{Ikeda:2007sa}
  M.~Ikeda {\it et al.}  [Super-Kamiokande Collaboration],
  %``Search for Supernova Neutrino Bursts at Super-Kamiokande,''
  Astrophys.\ J.\  {\bf 669} (2007) 519
  [arXiv:0706.2283 [astro-ph]].
  %%CITATION = ASJOA,669,519;%%


%SN spectrum
\bibitem{Totani:1997vj}
  T.~Totani, K.~Sato, H.~E.~Dalhed and J.~R.~Wilson,
  %``Future detection of supernova neutrino burst and explosion mechanism,''
  Astrophys.\ J.\  {\bf 496} (1998) 216
  [arXiv:astro-ph/9710203].
  %%CITATION = ASJOA,496,216;%%


\bibitem{Keil:2002in}
  M.~T.~Keil, G.~G.~Raffelt and H.~T.~Janka,
  %``Monte Carlo study of supernova neutrino spectra formation,''
  Astrophys.\ J.\  {\bf 590}, 971 (2003)  [arXiv:astro-ph/0208035].
  %%CITATION = ASTRO-PH 0208035;%%

\bibitem{snnew}
  S.~Hannestad and G.~Raffelt,
  %``Supernova neutrino opacity from nucleon nucleon bremsstrahlung and related
  %processes,''
  Astrophys.\ J.\  {\bf 507}, 339 (1998) [arXiv:astro-ph/9711132]; R.~Buras {\it
et al.},
%, H.~T.~Janka, M.~T.~Keil, G.~G.~Raffelt and M.~Rampp,
  %``Electron-neutrino pair annihilation: A new source for muon and tau
  %neutrinos in supernovae,''
  Astrophys.\ J.\  {\bf 587}, 320 (2003)
[arXiv:astro-ph/0205006]; M.~Liebendoerfer {\it et al.},
%, M.~Rampp, H.~T.~Janka and A.~Mezzacappa,
  %``Supernova simulations with Boltzmann neutrino transport: A comparison  of
  %methods,''
  Astrophys.\ J.\  {\bf 620}, 840 (2005) 
[arXiv:astro-ph/0310662]; G.~G.~Raffelt, M.~T.~Keil, R.~Buras, H.~T.~Janka and
M.~Rampp,
%``Supernova neutrinos: flavour-dependent fluxes and spectra,''
arXiv:astro-ph/0303226.
%%CITATION = ASTRO-PH 0303226;%%

\bibitem{msw1}
  L.~Wolfenstein,
  Phys.\ Rev.\ D {\bf 17}, 2369 (1978);

\bibitem{msw2}
  S.~P.~Mikheev and A.~Y.~Smirnov,
  Sov.\ J.\ Nucl.\ Phys.\  {\bf 42}, 913 (1985)
  [Yad.\ Fiz.\  {\bf 42}, 1441 (1985)];
%
  S.~P.~Mikheev and A.~Y.~Smirnov,
  Nuovo Cim.\ C {\bf 9}, 17 (1986).

\bibitem{msw3}
  V.~D.~Barger, K.~Whisnant, S.~Pakvasa and R.~J.~N.~Phillips,
  %``Matter effects on three-neutrino oscillations,''
  Phys.\ Rev.\ D {\bf 22}, 2718 (1980).

%global data

\bibitem{solglobal}
M.~Maltoni {\it et al.},
%T.~Schwetz, M.~A.~Tortola and J.~W.~F.~Valle,
%%``Status of global fits to neutrino oscillations,''
New J.\ Phys.\  {\bf 6}, 122 (2004) [hep-ph/0405172]; S.~Choubey,
%   ``Probing The Neutrino Mass Matrix In Next Generation Neutrino Oscillation
  %Experiments,''
  arXiv:hep-ph/0509217; S.~Goswami,
  %``Neutrino oscillations and masses,''
  Int.\ J.\ Mod.\ Phys.\ A {\bf 21}, 1901 (2006); A.~Bandyopadhyay {\it et al.},
%S.~Choubey, S.~Goswami, S.~T.~Petcov and D.~P.~Roy,
%   ``Update Of The Solar Neutrino Oscillation Analysis With The 766-Ty  Kamland Spectrum,''
  Phys.\ Lett.\ B {\bf 608}, 115 (2005)  [arXiv:hep-ph/0406328]; G.~L.~Fogli
{\it et al.},
%, E.~Lisi, A.~Marrone and A.~Palazzo,
  %``Global analysis of three-flavour neutrino masses and mixings,''
  Prog.\ Part.\ Nucl.\ Phys.\  {\bf 57}, 742 (2006)  [arXiv:hep-ph/0506083];

%Spectral splits
\bibitem{Spectral_split}
  H.~Duan, G.~M.~Fuller, J.~Carlson and Y.~Z.~Qian,
  %``Simulation of coherent non-linear neutrino flavor transformation in the
  %supernova environment. I: Correlated neutrino trajectories,''
  Phys.\ Rev.\  D {\bf 74} (2006) 105014
  [arXiv:astro-ph/0606616]; H.~Duan, G.~M.~Fuller, J.~Carlson and
Y.~Z.~Qian,
  %``Coherent development of neutrino flavor in the supernova environment,''
  Phys.\ Rev.\ Lett.\  {\bf 97} (2006) 241101
  [arXiv:astro-ph/0608050]; S.~Samuel,
  %``Neutrino oscillations in dense neutrino gases,''
  Phys.\ Rev.\  D {\bf 48} (1993) 1462; V.~A.~Kostelecky and S.~Samuel,
  %``Selfmaintained coherent oscillations in dense neutrino gases,''
  Phys.\ Rev.\  D {\bf 52} (1995) 621
  [arXiv:hep-ph/9506262]; J.~T.~Pantaleone,
  %``Stability Of Incoherence In An Isotropic Gas Of Oscillating Neutrinos,''
  Phys.\ Rev.\  D {\bf 58} (1998) 073002; S.~Samuel,
  %``Bimodal coherence in dense selfinteracting neutrino gases,''
  Phys.\ Rev.\  D {\bf 53} (1996) 5382
  [arXiv:hep-ph/9604341]; G.~G.~Raffelt and A.~Y.~Smirnov,
  %``Self-induced spectral splits in supernova neutrino fluxes,''
  Phys.\ Rev.\  D {\bf 76} (2007) 081301
  [Erratum-ibid.\  D {\bf 77} (2008) 029903]
  [arXiv:0705.1830 [hep-ph]]; G.~G.~Raffelt and A.~Y.~Smirnov,
  %``Adiabaticity and spectral splits in collective neutrino transformations,''
  Phys.\ Rev.\  D {\bf 76} (2007) 125008
  [arXiv:0709.4641 [hep-ph]]; B.~Dasgupta and A.~Dighe,
  %``Collective three-flavor oscillations of supernova neutrinos,''
  arXiv:0712.3798 [hep-ph].
  %%CITATION = ARXIV:0712.3798;%%
  %%CITATION = PHRVA,D52,621;%%
  %%CITATION = PHRVA,D48,1462;%%
  %%CITATION = PHRVA,D53,5382;%%
%%CITATION = PHRVA,D58,073002;%% 
  
%SK detector
\bibitem{raffeltforplusrev}
  R.~Tomas, M.~Kachelriess, G.~Raffelt, A.~Dighe, H.~T.~Janka and L.~Scheck,
  %``Neutrino signatures of supernova shock and reverse shock propagation,''
  JCAP {\bf 0409}, 015 (2004).
%  [arXiv:astro-ph/0407132].
  %%CITATION = ASTRO-PH 0407132;%

%sn1987a events
\bibitem{Loredo:2001rx}
  T.~J.~Loredo and D.~Q.~Lamb,
  %``Bayesian analysis of neutrinos observed from supernova SN 1987A,''
  Phys.\ Rev.\  D {\bf 65} (2002) 063002
  [arXiv:astro-ph/0107260].
  %%CITATION = PHRVA,D65,063002;%%
    
\bibitem{Tomas:2003xn}
  R.~Tomas, D.~Semikoz, G.~G.~Raffelt, M.~Kachelriess and A.~S.~Dighe,
  %``Supernova pointing with low- and high-energy neutrino detectors,''
  Phys.\ Rev.\ D {\bf 68} (2003) 093013 [arXiv:hep-ph/0307050].
%  [arXiv:hep-ph/0307050].
  %%CITATION = HEP-PH 0307050;%%
    
   %Other detectors
\bibitem{NeutrinoDetectors}
  A.~Bueno, I.~Gil-Botella and A.~Rubbia,
  %``Supernova neutrino detection in a liquid argon TPC,''
  arXiv:hep-ph/0307222; I.~Gil-Botella and A.~Rubbia,
  %``Oscillation effects on supernova neutrino rates and spectra and  detection
  %of the shock breakout in a liquid argon TPC,''
  JCAP {\bf 0310} (2003) 009
  [arXiv:hep-ph/0307244]; I.~Gil-Botella and A.~Rubbia, JCAP {\bf 0408} (2004)
001 [arXiv:hep-ph/0404151].
  %%CITATION = JCAPA,0408,001;%%


\bibitem{IEEE1588}
See, e.g., {\tt http://ieee1588.nist.gov/}.

\bibitem{OWGPS} M.~A.~Lombardi, L.~M.~Nelson, A.~N.~Novick and V.~S.~Zhang, 
{\it Time and Frequency Measurements Using the Global Positioning System}, 
Cal. Lab. Int. J. Metrology, (July-September, 2001) 26-33; 
see also: {\tt http://tf.nist.gov/time/oneway.htm};
{\tt http://tf.nist.gov/time/commonviewgps.htm}.

\bibitem{GPSDO} M.~A.~Lombardi and A.~N.~Novick, {\it Comparison of the one-way
and common view GPS measurement technique using a known frequency offset},
Proceedings of 34th Annual Precise Time and Time Interval (PTTI) Meeting;  
see also: {\tt http://tycho.usno.navy.mil/ptti/ptti2002/paper4.pdf}.

\bibitem{CPGPS} K.~Larson and J.~Levine, {\it Carrier-Phase Time Transfer},
IEEE Trans. On Ultrasonics, Ferroelectrics and Frequency Control, {\bf
46}, (1999) 1001; see also: {\tt http://tycho.usno.navy.mil/gpscp.html}.

\bibitem{ccf_grb} See, for example: D.L.~Band, Ap. J.
{\bf 486}, (1997) 928; J.~P.~Norris, G.~E.~Marani and J.~T.~Bonnell,
Ap. J. {\bf 534}, (2000) 340.

\end{thebibliography}
\end{document}